\documentclass[12pt]{article}
\usepackage{epsfig}
\usepackage{amssymb}
\usepackage{amsmath}
\usepackage{amsfonts}
\usepackage{graphicx}
\usepackage{mathrsfs}
\usepackage{mathabx}
\DeclareMathAlphabet{\mathscrbf}{OMS}{mdugm}{b}{n}
\usepackage[dvips]{color}
\usepackage{multirow}
\usepackage{calc}
\usepackage{accents}

\makeatletter
\newcommand{\shorteq}{\mathrel{\mkern0.2mu\mathpalette\shorteq@\relax\mkern0.2mu}}
\newcommand{\shorteq@}[2]{\scalebox{0.5}[1]{$\m@th#1=$}}

\newcommand{\longeq}[1]{\mathrel{\mathpalette\longeq@{#1}}}
\newcommand{\longeq@}[2]{%
  \begingroup
  \sbox\z@{$\m@th#1=$}%
  \ifdim#2<\wd\z@
    \resizebox{#2}{\height}{\box\z@}%
  \else
    \ifdim#2<3\wd\z@
      \hbox to #2{$\m@th#1=\hss=\hss=\hss=$}%
    \else
      \hbox to #2{$\m@th#1=\cleaders\hbox to 0.2\wd\z@{\hss$#1=$\hss}\hfil=$}%
    \fi
  \fi
  \endgroup
}
\makeatother

\newcommand{\vp}{{\vec p}}
\newcommand{\vq}{{\vec q}}

\newcommand{\bsigma}{\boldsymbol{\sigma}}

\newcommand{\bnabla}{\boldsymbol{\nabla}}

\newcommand{\bOmega}{\boldsymbol{\Omega}}

\newcommand{\R}{\mathbb{R}}
\newcommand{\C}{\mathbb{C}}

\newcommand{\fc}{\mathfrak{c}}

\newcommand{\ff}{\mathfrak{f}}

\newcommand{\fm}{\mathfrak{m}}
\newcommand{\fn}{{\mathfrak{n}}}
\newcommand{\ft}{\mathfrak{t}}

\newcommand{\fs}{\mathfrak{s}}

\newcommand{\fz}{\mathfrak{z}}

\newcommand{\fB}{\mathfrak{B}}

\newcommand{\fM}{\mathfrak{M}}

\newcommand{\fT}{\mathfrak{T}}

\newcommand{\bfa}{\mathbf{a}}

\newcommand{\bfe}{\mathbf{e}}

\newcommand{\bk}{\mathbf{k}}

\newcommand{\bn}{{\mathbf{n}}}
\newcommand{\bfr}{\mathbf{r}}
\newcommand{\bp}{\mathbf{p}}

\newcommand{\bu}{\mathbf{u}}

\newcommand{\bcA}{\boldsymbol{\cA}}

\newcommand{\bG}{\mathbf{G}}
\newcommand{\bH}{\mathbf{H}}
\newcommand{\bI}{\mathbf{I}}

\newcommand{\bM}{\mathbf{M}}

\newcommand{\bU}{\mathbf{U}}

\newcommand{\cA}{{\mathcal{A}}}

\newcommand{\cH}{\mathcal{H}}

\newcommand{\cF}{\mathcal{F}}

\newcommand{\cK}{\mathcal{K}}

\newcommand{\cN}{\mathcal{N}}
\newcommand{\cO}{\mathcal{O}}
\newcommand{\cP}{\mathcal{P}}

\newcommand{\cS}{\mathcal{S}}
\newcommand{\cT}{\mathcal{T}}
\newcommand{\cU}{\mathcal{U}}

\newcommand{\cW}{\mathcal{W}}
\newcommand{\cX}{\mathcal{X}}
\newcommand{\cY}{\mathcal{Y}}
\newcommand{\cZ}{\mathcal{Z}}
\newcommand{\be}{\begin{equation}}
\newcommand{\ee}{\end{equation}}
\newcommand{\bea}{\begin{eqnarray}}
\newcommand{\eea}{\end{eqnarray}}
\newcommand{\nn}{\nonumber}
\newcommand{\kt}{\rangle}
\newcommand{\br}{\langle}

\newcommand{\ed}{\end{document}}

\newcommand{\ra}{{\rm a}}
\newcommand{\rk}{{\rm k}}
\newcommand{\rn}{{\rm n}}
\newcommand{\ru}{{\rm u}}

\newcommand{\bi}{\begin{itemize}}
\newcommand{\ei}{\end{itemize}}

\newcommand{\bce}{\begin{center}}
\newcommand{\ece}{\end{center}}

\newcommand{\sD}{\mathscr{D}}

\newcommand{\sF}{\mathscr{F}}

\newcommand{\sH}{\mathscr{H}}

\newcommand{\sM}{\mathscr{M}}
\newcommand{\bsM}{\mathscrbf{M}}

\newcommand{\sR}{\mathscr{R}}

\newcommand{\sT}{\mathscr{T}}
\newcommand{\sV}{\mathscr{V}}

\newcommand{\sX}{\mathscr{X}}
\newcommand{\sZ}{\mathscr{Z}}

\newcommand{\bcK}{{\boldsymbol{\cK}}}

\newcommand\bfM{\boldsymbol{\fM}}

\newcommand{\bcH}{{\boldsymbol{\cH}}}
\newcommand{\bcU}{{\boldsymbol{\cU}}}

\newcommand{\bzero}{{\boldsymbol{0}}}

\newcommand{\for}{{\mbox{\rm for}}}

\newcommand{\sinc}{{\rm sinc}}

\newcommand{\mr}{\mathring}

\newcommand{\vk}{{\vec{\rk}}}

\newcommand{\cz}{{\check{z}}}

\begin{document}

\title{Acoustic scattering from quantum dynamics}

\author{Farhang Loran\thanks{E-mail address: loran@iut.ac.ir}~ and 
Ali~Mostafazadeh\thanks{Corresponding author, 
E-mail address: amostafazadeh@ku.edu.tr}\\[6pt]
$^*$Department of Physics, Isfahan University of Technology, \\ Isfahan 84156-83111, Iran\\[6pt]$^\dagger$
Departments of Mathematics and Physics, Ko\c{c} University,\\  34450 Sar{\i}yer,
Istanbul, T\"urkiye}

\date{ }
\maketitle

\begin{abstract}

Scattering of sound waves propagating in a compressible fluid is linked to the quantum dynamics generated by a class of non-Hermitian Hamiltonian operators. The scattering amplitude is expressed in terms of the corresponding evolution operator. This is used to provide a new proof of the reciprocity theorem in acoustic scattering, conduct analytic low-frequency acoustic scattering calculations, and study the scattering of sound waves by a finite planar array of delta-function point scatters embedded in an infinite acoustic slab. {A detailed discussion of acoustic scattering properties of a homogeneous acoustic slab including a pair of delta-function point scatters on one of its faces is provided.}
\end{abstract}

\section{Introduction} 

{Acoustic scattering is a physical phenomenon of great importance that has attracted much attention since the 19th century \cite{Rayleigh-book}.} The scattering of time-harmonic sound waves scattered by localized inhomogeneities of a compressible fluid is defined by Bergmann's equation  \cite{Bergmann,Pierce},	
	\be
	\rho\, \bnabla\cdot(\rho^{-1} \bnabla \psi)+k^2\fn^2\psi=0,
	\label{bergmann-eq-1}
	\ee
where $\psi$, $\rho$, and $\fn$ are respectively the pressure, density, and refractive index of the fluid, which are functions of the position vector $\bfr$, and $k$ is the incident wavenumber \cite{Martin}. The refractive index is related to the speed $c$ of sound in the fluid according to $\fn(\bfr):=c_0/c(\bfr)$, where $c_0$ is the speed of sound at spatial infinity. Eq.~\eqref{bergmann-eq-1} is equivalent to
	\be
	\bnabla\cdot(\rho^{-1} \bnabla \psi)+{c_0^2}k^2\kappa\,\psi=0,
	\label{bergmann-eq}
	\ee
where $\kappa(\bfr):={\fn(\bfr)^2/c_0^2\rho(\bfr)=[c(\bfr)^2\rho(\bfr)]^{-1}}$ is the compressibility of the fluid.

In regions of space in which $\nabla^2\rho$ exists, Bergmann's equation \eqref{bergmann-eq} is equivalent to the time-independent Schr\"odiger equation,
	\be
	(-\nabla^2+v)\phi=k^2\phi,
	\label{sch-eq}
	\ee
where  
	\begin{align}
	&\phi:=\rho^{-1/2}\psi,
	&&v :=k^2(1-\fn^2)+\rho^{1/2}\nabla^2\rho^{-1/2}.
	\label{v=}
	\end{align}
If the inhomogeneities of the medium are localized, there are constants $a$ and $\rho_0$ such that $\rho(\bfr)=\rho_0$, $\fn(\bfr)=1$, and $\kappa(\bfr)={\kappa_0:=(c_0^2\rho_0)^{-1}}$ for $r:=|\bfr|\geq a$. This implies that $\psi(\bfr)=\phi(\bfr)$ and $v(\bfr)=0$ for $r\geq a$. Consequently, solutions of \eqref{bergmann-eq} tend to superpositions of plane waves at spatial infinities. In particular, \eqref{bergmann-eq} admits scattering solutions satisfying
	\be
	\psi(\bfr)\to
	\cN_0\left[e^{i\bk_0\cdot\bfr}+\frac{e^{ikr}}{r}\,\ff(\hat\bn_0,\hat\bn)\right]~~~\for~~~r\to\infty,
	\label{scattering}
	\ee
where $\cN_0$ is the amplitude of the incident wave, $\bk_0$ is incident wave vector, $\ff$ is the scattering amplitude, and $\hat\bn_0$ and $\hat\bn$ are respectively the unit vectors along $\bk_0$ and $\bfr$,  \cite{Landau-Lifshitz-QM,Sakurai}, i.e., 
	\begin{align}
	&\hat\bn_0:=k^{-1}\bk_0,
	&&\hat\bn:=r^{-1}\bfr.
	\nn
	\end{align}
We arrive at the same conclusions for more general situations where the fluid inhomogeneities are not localized, but {the potential $v$ of Eq.~\eqref{v=} is an absolutely integrable function,} and $\rho(\bfr)-\rho_0$ and $\fn(\bfr)-1$ decay to zero as $r\to\infty$ at a rate that 
	\be
	\lim_{r\to\infty} r[\rho(\bfr)-\rho_0]=\lim_{r\to\infty} r^2 v(\bfr)=0.
	\label{condi-short}
	\ee 
Under these conditions, $v$ is a short-range potential \cite{Yafaev}, and \eqref{sch-eq} and consequently \eqref{bergmann-eq} admit scattering solutions fulfilling \eqref{scattering}. 

Because $\nabla^2\rho$ may {have singularities} for $r<a$, it is not generally possible to reduce the scattering problem defined by the Bergmann equation to that by the time-independent Schr\"odiger equation. There are, however, well-known Green's function methods similar to those of potential scattering that apply to the scattering phenomena modeled by the Bergmann equation \cite{Martin}.  

In Refs.~\cite{ap-2014,pra-2021}, we outline an alternative dynamical formulation of potential scattering in which the scattering phenomenon is linked to the dynamics of an effective non-unitary quantum system. More precisely, the evolution operator for the latter determines an integral operator, called the fundamental transfer matrix, which provides a tool for computing the scattering amplitude. This approach to potential scattering admits an extension to electromagnetic scattering \cite{pra-2023} in three dimensions and the scattering of transverse electric (TE) and transverse magnetic (TM) waves by inhomogeneities of effectively one- and two-dimensional isotropic media \cite{ptep-2024b,ptep-2026a}. The purpose of the present article is to extend this approach to acoustic scattering defined by Bergmann's equation \eqref{bergmann-eq} in three dimensions {and explore some of its applications.}

The basic motivation for the present study is the effectiveness of the dynamical formulation of stationary scattering in performing analytic scattering calculations. Specific examples are its utility in the derivation of explicit formulas for the scattering amplitude at low frequencies \cite{ptep-2026a,jpa-2021,jpa-2025} and the treatment of delta-function point scatterers \cite{pra-2021,pra-2023,ap-2022}. Other notable applications of this approach are the construction of the potentials and permittivity-permeability profiles for which the $N$-th order Born approximation is exact \cite{jpa-2024}, and a new proof of the reciprocity theorem in potential scattering \cite{ap-2026}. 

The outline of the remainder of this article is as follows. In Sec.~2, we {describe} the concept of fundamental transfer matrix and discuss its utility in determining the scattering amplitudes in potential and acoustic scattering. In Sec.~3, we derive the expression for the effective Hamiltonian operator whose evolution operator determines the acoustic fundamental transfer matrix. In Secs.~4 and 5, we use the properties of the effective Hamiltonian to provide an alternative proof of the {acoustic reciprocity theorem} and obtain analytic formulas for the leading- and next-to-leading-order terms in the low-frequency series expansion of the scattering amplitude, respectively. In Sec.~6, we apply our general results to study { the scattering of sound waves by an acoustic slab containing a planar array of delta-function point scatterers.} Here we obtain an explicit formula relating the scattering amplitude to the positions of point scatterers {and provide a detailed discussion of the scattering features of a slab with a pair of delta-function point scatterers on one of its faces. In Sec.~7, we present our concluding remarks.}

\section{Fundamental transfer matrix}

\begin{figure}
    \begin{center}
        \includegraphics[scale=.22]{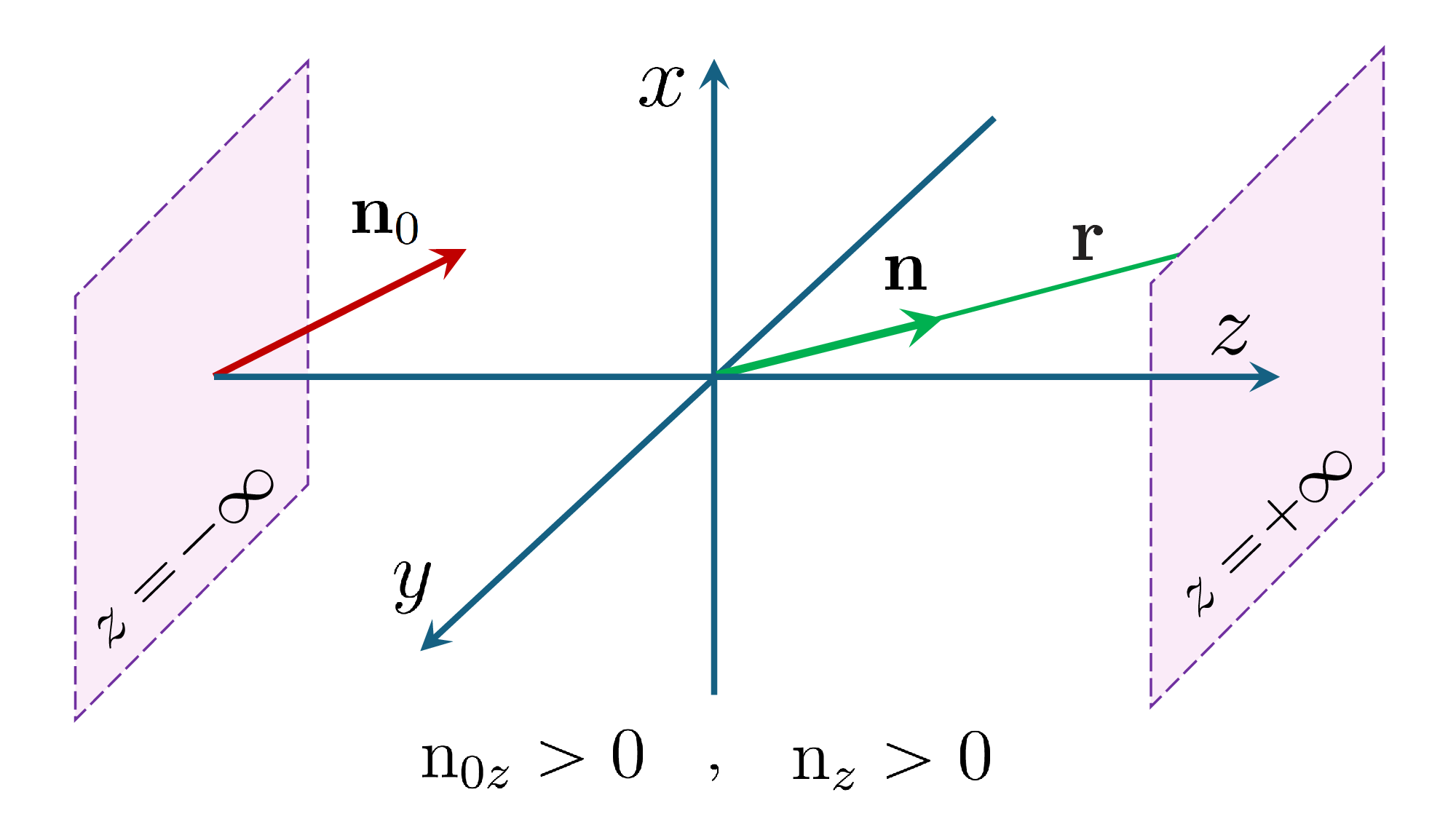}~~~~~~~
        \includegraphics[scale=.22]{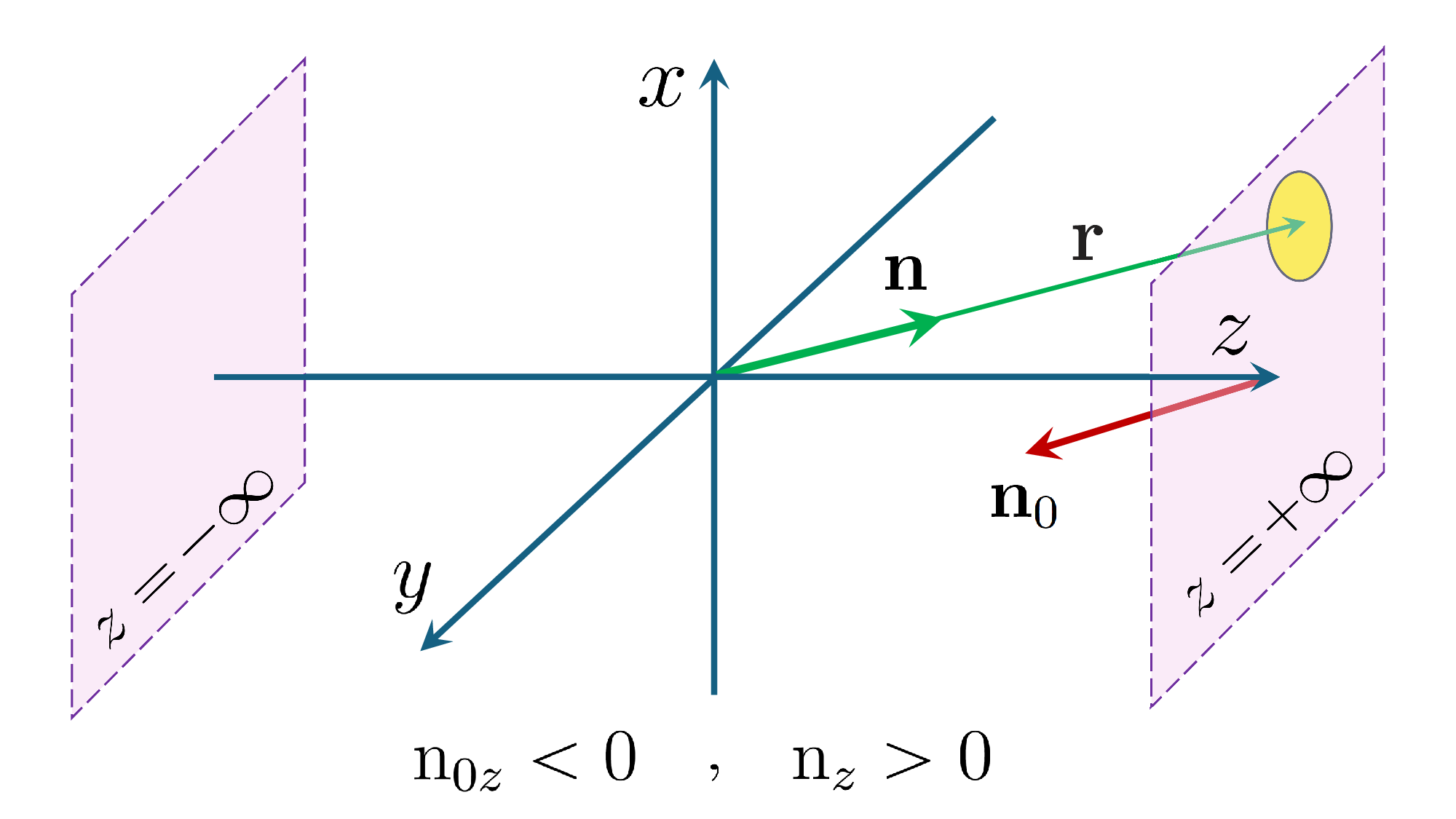}
        \caption{Schematic view of the scattering setup. $\bn_0$ and $\bn$ are respectively the unit vectors along the incident wave vector and the position vector for a generic detector. The yellow elliptic region represents a detector screen position at $z=+\infty$. This corresponds to $\bn$ having a positive $z$ component. The left and right panels represent situations in which the $z$ component of $\bn_0$ takes positive and negative values, respectively.}
        \label{fig1}
    \end{center}
    \end{figure}

Consider the scattering of a time-harmonic incident sound wave with angular frequency $\omega:=c_0/k$ and wave vector $\bk_0$.  Suppose that we choose a Cartesian coordinate system in which $\bk_0$ points towards the $x$-$y$ plane, and the detectors measuring the scattered wave are mounted on the planes corresponding to $z=\pm\infty$, as shown in Fig.~\ref{fig1}. 

Let $\vec r:=(x,y)$, so that $\bfr=(\vec r,z)$. More generally, for each $\bu\in\R^3$, {$\vec\ru$ denotes} $(\ru_x,\ru_y)$, where $\ru_x,\ru_y$, and $\ru_z$ are respectively the $x$, $y$, and $z$ components of $\bu$. This allows us to identify $\bu$ with $(\vec\ru,\ru_z)$. For example,  
	\begin{align}
	&\bk_0=(\rk_{0x},\rk_{0y},\rk_{0z})=(\vec\rk_0,\rk_{0z}),
	&&\vec\rk_0=(\rk_{0x},\rk_{0y}),\nn\\
	&\bn_0=(\rn_{0x},\rn_{0y},\rn_{0z})=k^{-1}(\vec \rk_0,\rk_{0z}),
	&&\bn=(\rn_x,\rn_y,\rn_z)=r^{-1}(\vec r,z).\nn
	\end{align} 
In the scattering setup we consider in this article, $\bk_0$ points towards the $x$-$y$ plane and the detectors lie in either of the planes $z=\infty$ and $z=-\infty$. Consequently,
	\begin{align}
	&\rn_{0z}\neq 0\neq \rn_{z}.
	\label{condi-ns}
	\end{align}
	
Suppose that $\rho(\bfr)-\rho_0$ and $\fn(\bfr)-1$ decay to zero as $r\to\infty$ at such a rate that \eqref{condi-short} holds and the potential $v$ of Eq.~\eqref{v=} is a short-range potential. Then, bounded solutions $\psi$ of \eqref{bergmann-eq} tend to superpositions of plane waves at spatial infinities. In view of \eqref{condi-ns}, we can express this condition in the form
	\be
	\psi(\vec r,z)\to
	\int_{\sD_k} \frac{d{\vec p}}{4\pi^2\varpi({\vec p})}\, e^{i{\vec p}\cdot{\vec r}}	\left[A_\pm({\vec p})\,e^{i\varpi({\vec p})z}+B_\pm({\vec p})\, e^{-i\varpi({\vec p})z}\right]~~\for~~z\to\pm\infty,
	\label{asymp}
	\ee
where $\vec p:=(p_x,p_y)$, ${\vec p}\cdot{\vec r}:=xp_x+yp_y$, 
	\begin{align}
	&\sD_k:=\big\{\vec p\in\R^2\,\big|\,|\vec p|<k\big\},
	&&\varpi({\vec p}):=\left\{\begin{array}{ccc}
	\sqrt{k^2-{\vec p}^{\,2}}&\for&{\vec p}\in\sD_k,\\ 
	i\sqrt{{\vec p}^{\,2}-k^2}&\for&{\vec p}\notin\sD_k,
	\end{array}\right.
    \label{varpi=}
	\end{align}
 and $A_\pm$ and $B_\pm$ are coefficient functions\footnote{\label{footnote-1}Throughout this article we use the term ``function'' to mean a ``tempered distribution'' \cite{strichartz}.} \cite{pra-2021}. 
 
 To view the integral in \eqref{asymp} as a Fourier integral, we identify $A_\pm$ and $B_\pm$ with functions defined on $\R^2$ that vanish outside the disc $\sD_k$, i.e.,  demand that $A_\pm,B_\pm\in\sF_k$, where
	\be
	\sF_k:=\big\{\,\phi\in\sF\,\big|\, \phi(\vec p\,)=0~\for~|\vec p\,|\geq k\,\big\},\nn
	\ee
 and $\sF$ is the set of functions $\phi:\R^2\to\C$.
 
 {Let $\sF^2$ and $\sF_k^2$ stand for the vector spaces of two-component functions with components belonging to $\sF$ and $\sF_k$, respectively. Denoting the space of $m\times n$ complex matrices by $\C^{m\times n}$, we have $\sF^2=\sF\otimes\C^{2\times 1}$ and $\sF_k^2=\sF_k\otimes\C^{2\times 1}$.} The fundamental transfer matrix \cite{pra-2021} is the linear operator $\widehat\bM:\sF_k^2\to\sF_k^2$ satisfying
 	\be 
	\left[\begin{array}{c}
	A_+ \\ B_+ \end{array}\right]=\widehat{\bM}\left[\begin{array}{c}
	A_- \\ B_- \end{array}\right].
	\label{M-def}
	\ee
This is a direct generalization of the defining condition for the transfer matrix $\bM$ of potential scattering in one dimension \cite{sanchez,tjp-2020}. Notice, however, that unlike the entries of $\bM$ which {are complex numbers,} the entries $\widehat M_{ab}$ of $\widehat\bM$ are linear operators acting in an infinite-dimensional function space, namely $\sF_k$. 

Similarly to $\bM$, the fundamental transfer matrix stores the information about the scattering properties of the system. More precisely, it determines the scattering amplitude $\ff$ of the fluid. Since, under the condition~\eqref{condi-short}, the asymptotic expressions for the solutions of the Bergmann and Schr\"odinger equations, \eqref{bergmann-eq} and \eqref{sch-eq}, have identical forms, the expression giving $\ff$ in terms of the entries of $\widehat\bM$ coincides with that found in potential scattering. As shown in \cite{ap-2026}, the latter has the form,
	\be
	\ff(\bn_0,\bn)=-2\pi i\,\varpi(\vec \rk_0)\times\left\{\begin{array}{ccc}
	\!\!\!\br\vec\rk|(\widehat M_{11}-\widehat M_{12}\widehat M_{22}^{-1}\widehat M_{21}-\widehat I)|\vec\rk_0\kt
	&\for&\rn_{0z}>0~{\rm and}~\rn_z>0,\\[6pt]
	-\,\br\vec\rk|\,\widehat M_{22}^{-1}\widehat M_{21}|\vec\rk_0\kt
	&\for&\rn_{0z}>0~{\rm and}~\rn_z<0,\\[6pt]
	\br\vec\rk|\,\widehat M_{12}\widehat M_{22}^{-1}|\vec\rk_0\kt
	&\for&\rn_{0z}<0~{\rm and}~\rn_z>0,\\[6pt]
	\br\vec\rk|\,(\widehat M_{22}^{-1}-\widehat I)|\vec\rk_0\kt
	&\for&\rn_{0z}<0~{\rm and}~\rn_z<0,
	\end{array}\right.
	\label{ff=}
	\ee
where $\vec\rk:=k\,\vec\rn=kr^{-1}\vec r$,
and we have employed Dirac's bra-ket notation. In particular, given  $\vec p,\vec p\,'\in\R^2$ and a linear operator $\widehat L:\sF\to\sF$, $\br \vec p\,|\widehat L|\vec p{\,'}\kt$ stands for the integral kernel of $\widehat L$ satisfying,
	\[\big(\widehat L\phi\big)(\vec p\,)=\br\vec p\,|\widehat L\phi\kt=
	\int_{\R^2}\!\! d^2 p\,'~\br\vec p\,|\widehat L|\vec p\,'\kt\, \br\vec p\,'|\phi\kt,\]
{where} $d^2 p\,'=dp'_xdp'_y$ and $\br\vec p\,'|\phi\kt=\phi(\vec p\,')$. 

Notice that Eq.~\eqref{ff=} gives the scattering amplitude provided that \eqref{condi-ns} holds. This does not restrict the generality of our analysis, because $\ff$ is a continuous function of {$\bn_0$ or $\bn$; whenever $\bn_0$ or $\bn$ violates \eqref{ff=}, there are unit vectors $\bn'_0$ and $\bn'$, respectively, in the vicinity of  $\bn_0$ and $\bn$ that satisfy \eqref{ff=}, and we can identify $\ff(\bn_0,\bn)$ with the limit of $\ff(\bn'_0,\bn')$ as $\bn'_0\to\bn_0$ and $\bn'\to\bn$.}

\section{Dynamical formulation of stationary acoustic scattering}

In potential scattering, the fundamental transfer matrix can be expressed in terms of the evolution operator for an effective quantum system defined by a non-Hermitian Hamiltonian operator \cite{ap-2014,pra-2021}. The same applies to the scattering of TE and TM waves by inhomogeneities in effectively one- and two-dimensional isotropic media, {which are governed by} Bergmann's equation in one and two dimensions \cite{ptep-2024b,ptep-2026a}. In the following, we follow the approach of \cite{ptep-2026a} to identify an effective Hamiltonian whose evolution operator gives the acoustic fundamental transfer matrix.

First, we introduce the scaled density and compressibility of the carrier fluid,
	\begin{align}
	&\alpha(\bfr):=\frac{\rho(\bfr)}{\rho_0},
	&\beta(\bfr):=\frac{\kappa(\bfr)}{\kappa_0}=\frac{c_0^2\rho_0}{c(\bfr)^2\rho(\bfr)}=
	\frac{\fn^2(\bfr)}{\alpha(\bfr)},
	\label{a-b-def}
	\end{align}
which satisfy
	\be
	\lim_{z\to\pm\infty}\alpha(\vec r,z)=\lim_{z\to\pm\infty}\beta(\vec r,z)=1.
	\label{ab-limits}
	\ee
In terms of $\alpha$ and $\beta$, Bergmann's equation~\eqref{bergmann-eq} takes the form
	\be
	\bnabla\cdot(\alpha^{-1} \bnabla \psi)+k^2\beta\,\psi=0.
	\label{bergmann-eq-ab}
	\ee

Next, for each real number $z$, we let $\psi(z):\R^2\to\C$ be the function defined by $\big(\psi(z)\big)(\vec r):=\psi(\vec r,z)$, and $\tilde\psi(z):\R^2\to\C$ be its Fourier transform,  i.e., for all $\vec p\in\R^2$,
	\be
	\big(\tilde\psi(z)\big)(\vec p):=\tilde\psi(\vec p,z):=\int_{\R^2}d^2\vec r\:e^{-i\vec p\cdot\vec r}\psi(\vec r,z)=\int_{-\infty}^\infty\!\! dx\int_{-\infty}^\infty\!\! dy\:
	e^{-i(x p_x+yp_y)}\psi(x,y,z).
	\label{FT}
	\ee
In view of 	\eqref{asymp}, \eqref{bergmann-eq-ab}, and \eqref{FT}, $\tilde\psi(z)$ defines a one-parameter family of elements of $\sF$ that satisfies  
	\begin{align}
	&-\partial_z\big[\widehat\alpha(z)^{-1}\partial_z\tilde\psi(z)\big]+\widehat\sV(z)\tilde\psi(z)=\widehat\varpi^2\tilde\psi(z),
	\label{tbergmann-eq}
	\end{align}
and
	\begin{align}
	&\tilde\psi(z)\to \widehat\varpi^{-1}\left(e^{iz\widehat\varpi}A_{\pm}+
	e^{-iz\widehat\varpi}B_{\pm}\right)~~~\for~~~z\to\pm\infty,
	\label{tasymp}
	\end{align}
where
	\begin{align}
	&\widehat\alpha(z):=\alpha(\widehat{\vec r},z)=\alpha(\widehat x,\widehat y,z),
	&&\widehat\sV(z):=\widehat{\vec p}\cdot\big[\widehat\alpha(z)^{-1}-\widehat I\,\big]
	\widehat{\vec p}+k^2\big[\widehat I-\widehat\beta(z)\big],
	\label{eq1}\\
	&\widehat\beta(z):=\beta(\widehat{\vec r},z)=\beta(\widehat x,\widehat y,z),
	&&\widehat\varpi:=\varpi(\widehat{\vec p}):=\sqrt{k^2\widehat I-(\widehat p_x^2+\widehat p_y^2)},
	\label{eq2}
	\end{align}
$\widehat{\vec r}:=(\widehat x,\widehat y)$ and $\widehat{\vec p}:=(\widehat p_x, \widehat p_y)$ are respectively the standard position and momentum operators of quantum mechanics in two dimensions, and $\widehat I$ is the identity operator.\footnote{We use the same symbol $\widehat I$ to denote the identity operators acting in $\sF$, $\sF_k$, $\sF^2$, and $\sF_k^2$. The same holds for the zero operators acting in these vector space which we denote by $\widehat 0$.} $\widehat{\vec r}$ and $\widehat{\vec p}$ act on elements $\phi$ of $\sF$ (and $\sF^2$) according to
	\begin{align}
	&\big(\,\widehat{\vec r}\,\phi\big)(\vec p)=i\vec\nabla_{\!p}\phi(\vec p)=
	\big(i\partial_{p_x}\phi(\vec p),i\partial_{p_y}\phi(\vec p)\big),
	&&\big(\,\widehat{\vec p}\,\phi\big)(\vec p)=\phi(\vec p)\vec p.
	\end{align}
In particular, 
	\be
	\big(\widehat\varpi\phi\big)(\vec p)=\varpi(\vec p)\phi(\vec p).
	\label{hat-varpi}
	\ee
	
Notice also that, given a function $g:\R^3\to\C$, we can identify $\widehat g(z):=g(\widehat{\vec r},z)$ with the integral operator satisfying \cite{pra-2021}:
	\be
	\big(\widehat g(z)\phi\big)(\vec p)=\frac{1}{4\pi^2}
	\int_{\R^2}d^2p\,'\:\tilde g(\vec p-\vec p\,',z)\phi(\vec p\,').
	\label{g-id}
	\ee
In particular, $\widehat\alpha(z), \widehat\beta(z)$, and $\widehat\sV(z)$ are integral operators acting in $\sF$ (and $\sF^2$). 

Next, consider the two-component wave function,
	\be
	\Psi(z):=\frac{1}{2}\,e^{-iz\widehat\varpi_r \widehat\bsigma_3}
	\left[\begin{array}{c} \widehat\varpi\,\tilde\psi(z)-i\widehat{\alpha}(z)^{-1}\,\partial_z\tilde\psi(z)\\
	\widehat\varpi\,\tilde\psi(z)+ i\widehat{\alpha}(z)^{-1}\,\partial_z\tilde\psi(z)\end{array}\right]\in\sF^2,
	\label{2-component}
	\ee
where $\widehat\bsigma_j$ stands for the linear operator that multiplies two-component functions  by the Pauli matrix $\bsigma_j$ for $j\in\{1,2,3\}$, i.e., 
	\be
	\big(\widehat\bsigma_j\phi\big)(\vec p):=\bsigma_j\phi(\vec p),
	\label{Pauli-op}
	\ee
and introduce
	\begin{align}
	&\widehat\varpi_r:=\widehat\Pi_k\widehat\varpi=\widehat\varpi\widehat\Pi_k,
	&& (\widehat\Pi_k\phi)(p):=\left\{\begin{array}{ccc}
	\phi(p)&\for&|p|<k,\\
	0&\for&|p|\geq k.\end{array}\right.
	\label{project}
	\end{align}
In view of \eqref{ab-limits}, \eqref{tasymp}, and \eqref{2-component},
	\be
	\Psi(\pm\infty):=\lim_{z\to\pm\infty}\Psi(z)=\left[\begin{array}{c}
	A_\pm\\
	B_\pm\end{array}\right]\in\sF^2_k.
	\nn
	\ee
This relation allows us to write \eqref{M-def} as 
	\be
	\widehat\bM\,\Psi(-\infty)=\Psi(+\infty).
	\label{M-Psi}
	\ee
	
It is easy to show that \eqref{tbergmann-eq} is equivalent to the time-dependent Schr\"odinger equation,
	\be
	i\partial_z\Psi(z)=\widehat\bH(z)\Psi(z),
	\label{sch-eq-t}
	\ee
with $z$ playing the role of time, and $\widehat\bH(z)$ a ``time-dependent'' Hamiltonian operator of the form:
	\begin{align}
	&\widehat\bH(x):=\frac{1}{2}e^{-ix\widehat\varpi_r\widehat\bsigma_3} 
	\left\{ \widehat\sV(z)\widehat\varpi^{-1}\widehat\bcK
	+\widehat\varpi[\widehat I-\widehat\alpha(z)]\widehat\bcK^T\right\}
	 e^{ix\widehat\varpi_r\widehat\bsigma_3}- i\widehat\varpi_i\widehat\bsigma_3,
	\label{bH=}
	\end{align}
where $\widehat\bcK,\widehat\bcK^T:\sF^2\to\sF^2$ are the linear operators defined by  
	\begin{align}
	&\big(\widehat\bcK\Phi\big)(\vec p):=\bcK\Phi(\vec p),
	&&\big(\widehat\bcK^T\Phi\big)(\vec p):=\bcK^T\Phi(\vec p),
	&&\bcK:=\left[\begin{array}{cc}
	1 & 1\\
	-1 & -1\end{array}\right],
	\label{cK=}
	\end{align}
$\bcK^T$ is the transpose of $\bcK$, and 
	\be
	\widehat\varpi_i:=-i(\widehat\varpi-\widehat\varpi_r)=
	-i\widehat\varpi(\widehat I-\widehat\Pi_k).
	\label{w-i}
	\ee
Observe that we have extended the definitions of $\widehat\varpi, \widehat\varpi_r, \widehat\sV(z)$, and $\widehat\alpha(z)$ to $\sF^2$; they now apply to components of elements of $\sF^2$.
 
If we identify $\sF$ with $L^2(\R^2)$, we can view $\widehat\bH(z)$ as a linear operator acting in the Hilbert space $L^2(\R^2)\otimes\C^{2\times 1}$. In this way, we can speak of the adjoint of $\widehat\bH(z)$ and show that it is not a Hermitian operator. This implies that the dynamics generated by this operator in $L^2(\R^2)\otimes\C^{2\times 1}$ is not unitary. Denoting the corresponding evolution operator by $\widehat\bU(z,z_0)$, which satisfies $\widehat\bU(z,z_0)\Psi(z_0)=\Psi(z)$, 
we can use \eqref{M-Psi} to infer 
	\be
	\widehat\bM=\widehat\bU(+\infty,-\infty).
	\label{M=1}
	\ee
Repeating the analysis of Secs.~4 and 5 of Ref.~\cite{pra-2021}, we can also show that
	\be
	\widehat\bM=\widehat\Pi_k\,\widehat\bcU(+\infty,-\infty)\widehat\Pi_k,
	\label{M=2}
	\ee
where $\widehat\bcU(z,z_0)$ is the evolution operator for the Hamiltonian operator:
	\be
	\widehat\bcH(z):=\frac{1}{2}e^{-iz\widehat\varpi\bsigma_3} 
	\left\{\widehat\sV(z)\widehat\varpi^{-1}\bcK
	+\widehat\varpi[\hat I-\widehat\alpha(z)]\bcK^T\right\}
	 e^{iz\widehat\varpi \bsigma_3}.
	\label{bcH=}
	\ee

If the inhomogeneities of the fluid are confined to the region of space determined by $a_-\leq z\leq a_+$ for some $a_\pm\in\R$, then $\alpha(\vec r,z)=\beta(\vec r,z)=1$ for $z<a_-$ and $z>a_+$. According to  \eqref{eq1}, \eqref{eq2}, and \eqref{bcH=}, for these values of $z$, $\widehat\alpha(z)=\widehat\beta(z)=\widehat I$, $\widehat\sV(z)=\widehat 0$, and $\widehat\bcH(z)=\widehat 0$, where $\widehat 0$ is the zero operator. As a result, $\widehat\bcU(a_-,z)=\widehat I$ for $z\leq a_-$, and $\widehat\bcU(z,a_+)=\widehat I$ for $z\geq a_+$. Therefore, $\widehat\bcU(+\infty,-\infty)=\widehat\bcU(a_+,a_1)$, and  \eqref{M=2} becomes
	\be
	\widehat\bM=\widehat\Pi_k\,\widehat\bcU(a_+,a_-)\widehat\Pi_k.
	\label{M-local}
	\ee

\section{Reciprocity theorem in acoustic scattering} 

Acoustic reciprocity is among the most celebrated achievements of the study of sound waves in the 19th century. It was originally formulated in a precise manner by Lord Rayleigh \cite{Rayleigh-1873} in 1873 and has since been a focus of attention \cite{Fokkema,Achenbach,Masoud-2019}. It means that swapping the roles of the source and detector does not affect the outcome of a scattering experiment. More precisely, the scattering amplitude $\ff$ satisfies 
	\be
	\ff(\bn_0,\bn)=\ff(-\bn,-\bn_0).
	\label{reciprocity}
	\ee
	
Standard proofs of the reciprocity theorem for waves satisfying Helmholtz or time-independent Schr\"odinger equations make use of Green's identities \cite{Masoud-2019,Twersky1954,Bilhorn-1964}, time-reversal symmetry, or related properties of the relevant Green's functions \cite{Carminati,Deak-Fulop}. In Ref.~\cite{ap-2026}, we give an alternative proof of the reciprocity theorem in potential scattering which is based on the dynamical formulation of the potential scattering \cite{pra-2021}. In this section, we use this approach to provide a proof of the reciprocity theorem for acoustics scattering defined by Bergmann's equation \eqref{bergmann-eq-1}.

{Since the expression for the scattering amplitude in terms of the fundamental transfer matrix entries -- namely \eqref{ff=} -- is identical to its counterpart in potential scattering, we can utilize the results of Ref.~\cite{ap-2026} that stem from this expression. Specific examples include Theorems~3 and 5 of that reference, whose proofs follow directly from \eqref{ff=}. Consequently, their conclusions also apply to acoustic scattering problems governed by Bergmann's equation.} These theorems establish the equivalence of the reciprocity condition \eqref{reciprocity} and the following operator identity for the fundamental transfer matrix.
	\be
	(\widehat\bOmega\,\widehat\fT)^{-1}\widehat\bM^\dagger
(\widehat\bOmega\,\widehat\fT)\,\widehat\bM=\widehat I.
	\label{rec-1}
	\ee
Before giving the definitions of $\widehat\bOmega$ and $\widehat\fT$, we wish to point out that we respectively identify the function spaces $\sF$ and $\sF_k$ with the Hilbert spaces $L^2(\R^2)$ and $L^2(\sD_k)$. The operators $\widehat\bOmega$ and $\widehat\fT$ are linear operators acting in $\sF^2=L^2(\R^2)\otimes\C^{2\times 1}$ according to
	\begin{align}
	&\big(\widehat\bOmega\Phi\big)(\vec p):=
	\left[\begin{array}{cc}
	0 & 1\\
	-1 &0\end{array}\right]\Phi(\vec p), 
	&&\widehat\fT:=\widehat\varpi^{-1}\widehat\cP\widehat\cT,
	\label{ops1}
	\end{align}
where
	\begin{align}
	&\big(\widehat\cP\Phi\big)(\vec p):=\Phi(-\vec p),
	&&\big(\widehat\cT\Phi\big)(\vec p):=\Phi(\vec p)^*.
	\label{ops2}
	\end{align}
	
It is important to note that $\widehat\bOmega,\widehat\cP,\widehat\cT$, and consequently $\widehat\fT$ leave $\sF_k^2=L^2(\sD_k)\otimes\C^{2\times 1}$ invariant. Therefore, \eqref{rec-1} is an operator identity in $L^2(\sD_k)\otimes\C^{2\times 1}$, and $\widehat\bM^\dagger$ stands for the adjoint of $\widehat\bM$ which is a linear operator acting in this Hilbert space.
The following is a list of basic properties of $\widehat\bOmega,\widehat\cP$, $\widehat\cT$, and $\widehat\fT$.
	\begin{enumerate}
	\item $\widehat\bOmega,\widehat\cP$, and $\widehat\cT$ are mutually commuting operators that  satisfy
	\begin{align}
	&\widehat\bOmega^2=-\widehat\cP^2=-\widehat\cT^2=-\widehat I,
	\label{id6}\\
	&\widehat\bOmega\;\widehat{\vec p}\;\widehat\bOmega^{-1}=
	-\widehat\cP\;\widehat{\vec p}\;\widehat\cP^{-1}=
	\widehat\cT\;\widehat{\vec p}\;\widehat\cT^{-1}=\widehat{\vec p}.
	\label{id16}
	\end{align}
	\item $\widehat\bOmega$ and $\widehat\cP$ are linear operators, while $\widehat\cT$ and $\widehat\fT$ are antilinear operators. 
	\item $\widehat\bOmega$ is an anti-Hermitian operator, while 
	$\widehat\cP$, $\widehat\cT$, and $\widehat\fT$ are Hermitian operators.
	\end{enumerate}
	
Next, we introduce 
	\be
	\widehat\sT:=\widehat\cP\widehat\cT,
	\label{sT}
	\ee 
so that $\widehat\fT=\widehat\varpi^{-1}\widehat\sT$. Then, $\widehat\sT$ is a Hermitian antilinear operator satisfying $\widehat\sT\,^2=\widehat I$, and for every $\Phi\in L^2(\R)\otimes\C^{2\times 1}$ and $\vec r\in\R^2$,  
	\begin{align}
	\big(\widehat\sT\Phi\big)(\vec r\,)=
	\br\vec r\,|\widehat\sT|\Phi\kt&=\int_{\R^2}d^2\vec p\:\br\vec r\,|\vec p\,\kt
	\br\vec p\,|\widehat\cP\widehat\cT|\Phi\kt=\frac{1}{2\pi}\int_{\R^2}d^2\vec p\: e^{i\vec r\cdot\vec p}\,\Phi(-\vec p)^*\nn\\
	&=\frac{1}{2\pi}\int_{\R^2}d^2\vec p\,'\: e^{-i\vec r\cdot\vec p\,'}\Phi(\vec p\,')^*
	=\left[\frac{1}{2\pi}\int_{\R^2}d^2\vec p\,'\: e^{i\vec r\cdot\vec p\,'}\Phi(\vec p\,')\right]^*\nn\\
	&=\left[\int_{\R^2}d^2\vec p\,'\:\br\vec r\,|\vec p\,'\kt
	\br\vec p\,'|\Phi\kt\right]^*=\br\vec r\,|\Phi\kt^*=\Phi(\vec r\,)^*.\nn
	\end{align}
This calculation identifies $\widehat\sT$ with the operation of complex-conjugation in the position representation. We can use this observation to show that, given a function $g:\R^3\to\C$, the operator $\widehat g(z):=g(\widehat{\vec r},z)$ satisfies
	\be
	\widehat g(z)^\dagger=\widehat\sT\,\widehat g(z)\widehat\sT^{-1}.
	\label{id11}
	\ee
	
The following are consequences of \eqref{eq1}, \eqref{eq2}, \eqref{project}, \eqref{cK=}, \eqref{w-i}, and \eqref{ops1} -- \eqref{sT}.
	\begin{align}
	&\widehat\varpi_r^\dagger=\widehat\varpi_r=
	\widehat\cT\,\widehat\varpi_r\,\widehat\cT=
	\widehat\cP\,\widehat\varpi_r\,\widehat\cP=
	\widehat\sT\,\widehat\varpi_r\,\widehat\sT^{-1},
	\label{eq-zz1}\\
	&\widehat\varpi_i^\dagger=\widehat\varpi_i=
	\widehat\cT\,\widehat\varpi_i\,\widehat\cT=
	\widehat\cP\,\widehat\varpi_i\,\widehat\cP=
	\widehat\sT\,\widehat\varpi_i\,\widehat\sT^{-1},
	\label{eq-zz2}\\
	&\widehat\varpi^\dagger=\widehat\cT\,\widehat\varpi\,\widehat\cT=
	\widehat\cT\,\widehat\cP\,\widehat\varpi\,\widehat\cP\,\widehat\cT=
	\widehat\sT\,\widehat\varpi\,\widehat\sT^{-1},
	\label{eq-zz3}\\
	&[\widehat\varpi_r,\widehat\bOmega]=[\widehat\varpi_i,\widehat\bOmega]=
	[\widehat\varpi,\widehat\bOmega]=[\widehat{\sV}(x),\widehat\bOmega]=\widehat 0,
	\label{eq-zz45}\\
	&[\widehat\bcK,\widehat\cT\,]=[\widehat\bcK,\widehat\cP\,]=[\widehat\bcK,\widehat\sT\,]=\widehat 0,
	\label{eq-zz46}\\
	&\widehat\bOmega\,\widehat\bcK\,\widehat\bOmega^{-1}=-\widehat\bcK^T=-\widehat\bcK^\dagger,
	\label{eq-zz5}\\
	&\widehat\bOmega\,\widehat\bsigma_3\,\widehat\bOmega^{-1}=-\widehat\bsigma_3=-\widehat\bsigma_3^\dagger.
	\label{eq-zz6}
	\end{align}
		
Next, we recall the statement of Lemma~1 of Ref.~\cite{ap-2026}.
	\begin{itemize}
	\item[] {\bf Lemma 1:} Let $\widehat H(t)$ be a possibly time-dependent Hamiltonian operator acting in a Hilbert space $\sH$, $\widehat\bU(t,t_0)$ be the corresponding evolution operator, and $\widehat\cX:\sH\to\sH$ be a time-independent invertible antilinear operator. If $\widehat H(t)$  satisfies
	\be
	\widehat H(t)^\dagger=-\widehat\cX\,\widehat H(t)\,\widehat\cX^{-1},
	\label{ph-H}
	\ee
then $\widehat U(t,t_0)$ is $\cX$-anti-pseudo-unitary, i.e.,
	$\widehat U(t,t_0)^\dagger=\widehat\cX\,\widehat U(t,t_0)^{-1} \widehat\cX^{-1}$. In other words \eqref{ph-H} implies
	\be
	\widehat\cX^{-1}\widehat U(t,t_0)^\dagger\widehat\cX\,\widehat U(t,t_0)=\widehat I.
	\label{lemma1}
	\ee
	\end{itemize} 
	\begin{itemize}
	\item[] {\bf Proof of Reciprocity Theorem:} Because the reciprocity condition \eqref{reciprocity} is equivalent to \eqref{rec-1}, it is sufficient to show that the latter holds. To do this, first, we use \eqref{eq1}, \eqref{eq2}, and \eqref{id11} -- \eqref{eq-zz3} to establish:
	\be
	\widehat{\sV}(x)^\dagger=\widehat\sT\,\widehat{\sV}(x)\,\widehat\sT^{-1}.
	\label{eq-zz4}
	\ee
Using this equation together with \eqref{eq1}, \eqref{eq2}, \eqref{project}, \eqref{bH=} -- \eqref{w-i}, \eqref{ops1}, \eqref{ops2}, \eqref{sT}, and \eqref{eq-zz1} -- \eqref{eq-zz6}, we can show that
	\be
	\widehat\bOmega\,\widehat\fT\,\widehat\bH(z)(\widehat\bOmega\,\widehat\fT)^{-1}=-\widehat\bH(z)^\dagger.
	\label{H-anti-ph}
	\ee
In view of this relation and the fact that $\widehat\bOmega\,\widehat\fT$ is an invertible antilinear operator, the Hamiltonian operator $\widehat\bH(z)$ satisfies the hypothesis of Lemma~1. Therefore, the corresponding evolution operator, namely $\widehat\bU(z,z_0)$, fulfills
	\be
(\widehat\bOmega\,\widehat\fT)^{-1}\widehat\bU(z,z_0)^\dagger
(\widehat\bOmega\,\widehat\fT)\,\widehat\bU(z,z_0)=\widehat I.
	\label{ps-unitary1-bU}
	\ee
In view of \eqref{M=1} and \eqref{ps-unitary1-bU}, $\widehat\bM$ satisfies \eqref{rec-1}.~~~{$\Box$}
\end{itemize}

We close this section by noting that according to Eq.~\eqref{ff=}, the reciprocity condition \eqref{reciprocity} and the identity $\varpi(-\vec k)=\varpi(\vec k)$ imply
	\be
	f(\bn_0,\bn)=-2\pi i\,\varpi(\vec k)\,\br -\,\vec k_0|(\widehat M_{22}^{-1}-\widehat I\,)|\!-\vec k\kt~~~~\for~~~\rn_{0z}>0~~{\rm and}~~\rn_z>0.
	\label{f-left=}
	\ee

\section{Low-frequency acoustic scattering}

Suppose that the inhomogeneities of the fluid are confined to the region defined by $0\leq z\leq \ell$ for some positive real parameter $\ell$. We wish to explore the behavior of the scattering amplitude for incident wavenumbers $k$ that are much smaller than $\ell^{-1}$, i.e., $k\ell\ll 1$. This defines a notion of ``low-frequency scattering'' which is less restrictive than the standard definition of this term, as the latter requires the inhomogeneities of the fluid to be confined to a ball of radius $\ell$ and $k\ell\ll 1$. It is also worth mentioning that most of the known results on low-frequency acoustic scattering apply to waves satisfying the Helmholtz equation \cite{Kriegsmann-1983,dassios-book}, i.e., the career fluid has a constant mass density.

In Ref.~\cite{ptep-2026a}, we study the scattering of low-frequency TE and TM waves by an effectively two-dimensional medium. Because these waves satisfy Bergmann's equation in two dimensions, we can use the same analysis to study low-frequency acoustic scattering in three dimensions. To avoid redundancy, we skip the details of this analysis which are almost identical to those given in Ref.~\cite{ptep-2026a} for two dimensions, and present its final results.

As we show in Ref.~\cite{ptep-2026a}, the structure of the effective Hamiltonian operator \eqref{bcH=} allows for the expansion of the fundamental transfer matrix $\widehat\bM$ and its entries $\widehat M_{ab}$ in nonnegative integer powers of $k\ell$ and leads to a systematic iterative procedure for computing the coefficients of the corresponding series expansions. Substituting the latter in \eqref{ff=}, we arrive at a power series expansion of the scattering amplitude in positive integer powers of $k\ell$, i.e.,
	\be
	\ff(\bn_0,\bn)=\sum_{n=1}^\infty \ff^{(n)}(\bn_0,\bn)\,(k\ell)^n,
	\label{f-series}
	\ee
where $\ff^{(n)}$ are coefficient functions that do not depend on $k\ell$. 
{In particular, through a straightforward though lengthy calculation, we obtain explicit analytical expressions for the leading- and next-to-leading-order terms in \eqref{f-series}.} Before presenting these, we make a technical comment and introduce some auxiliary variables that enter the calculations of $\ff^{(1)}$ and $\ff^{(2)}$. 

As we noted above, our analysis of low-frequency scattering rests on the assumption that the inhomogeneities of the fluid are {confined to} the region defined by $0\leq z\leq\ell$. Furthermore, we suppose that the quantities $\alpha$ and $\beta$ that respectively represent the density and compressibility of the fluid are functions of $x,y$, and $z/\ell$. In other words, there are functions $w_\alpha,w_\beta:\R^2\times[0,1]\to\C$ such that 
	\begin{align}
	\alpha(\vec r,z)&:=\frac{\rho(\vec r,z)}{\rho_0}=1+\left\{\begin{array}{cc}
	w_\alpha(\vec r,z/\ell)&\for~~0\leq z\leq\ell,\\
	0&{\rm otherwise},\end{array}\right.
	\label{alpha=}\\
	\beta(\vec r,z)&:={\frac{\kappa(\vec r,z)}{\kappa_0}}=1+\left\{\begin{array}{cc}
	w_\beta(\vec r,z/\ell)&\for~~0\leq z\leq\ell,\\
	0&{\rm otherwise}.\end{array}\right.
	\label{beta=}
	\end{align}

Let, for each $z\in\R$, $\check z$ denote $z/\ell$, represent the unit vectors $\bn_0$ and $\bn$ by the corresponding polar and azimuthal angles $(\vartheta_0,\varphi_0)$ and $(\vartheta,\varphi)$ in the spherical coordinate system corresponding to the Cartesian coordinate system {with coordinates $(x,y,z)$}, and introduce 
	\begin{align}
	&v_\alpha(\vec r,\check z):=\frac{w_\alpha(\vec r,\check z)}{w_\alpha(\vec r,\check z)+1},
	&&\fc_0:=\cos\vartheta_0,
	&&\fc:=\cos\vartheta.
	\label{v-alpha=}
	\end{align}
Then, following the procedure outlined in Ref.~\cite{ptep-2026a}, we can express $\ff^{(1)}$ and $\ff^{(2)}$ in the form\footnote{{The two-dimensional analogs of \eqref{f1=} and \eqref{f2=} that are given in Ref.~\cite{ptep-2026a} differ from these equation in the numerical coefficients of some of the terms. This stems from the convention adopted in \cite{ptep-2026a} for the definition of the scattering amplitude in two dimensions.}},
	\begin{align}
	\ff^{(1)}(\bn_0,\bn)=&\frac{k}{4\pi}
	\Big[W_0(\vk,\vk_0)+\fc_0\fc \,
	\overline{\tilde w}_{\alpha,0}\big(\vk-\vk_0\big)\Big],
	\label{f1=}\\
	\ff^{(2)}(\bn_0,\bn)=&\frac{ik}{4\pi}\Bigg\{
	\fc_0\Big[X_1(\vk,\vk_0)+X_2\big(\vk-\vk_0\big)+W_1(\vk,\vk_0)
	-\fc^2\overline{\tilde w}_{\alpha,1}\big(\vk-\vk_0\big)\Big]\nn\\
	&\hspace{1.8cm}+\fc\Big[Y_1(\vk,\vk_0)+Y_2\big(\vk-\vk_0\big)-W_1(\vk,\vk_0)
	+\fc_0^2\overline{\tilde w}_{\alpha,1}\big(\vk-\vk_0\big)\Big]\nn\\
	&\hspace{1.8cm}+\frac{1}{8\pi^2}\int_{\sD_k}d^2\vq\left(1-q^2/k^2\right)^{-1/2}
	\Big[W_0(\vk,\vq)-\fc\left(1-q^2/k^2\right)^{1/2}\,\overline{\tilde w}_{\alpha,0}
	\big(\vk-\vq\big)\Big]
	\times\nn\\
	&\hspace{4.2cm}\Big[W_0(\vq,\vk_0)
	-\fc_0\left(1-q^2/k^2\right)^{1/2}\,\overline{\tilde w}_{\alpha,0}\big(\vq-\vk_0\big)\Big]\Bigg\},
	\label{f2=}
	\end{align}
where, for each $l\in\{0,1\}$,
	\begin{align}
	& W_l(\vp,{\vp\,'}):=k^{-2}\vp\,\cdot{\vp\,'}\:\overline{\tilde v}_{\alpha,l}(\vp-{\vp\,'})+\overline{\tilde w}_{\beta,l}(\vp-{\vp\,'}),
	\label{W-def}\\	
	&\overline{\tilde w}_{\alpha,l}(\vp):=\int_0^1d\cz\:\cz^l\:\tilde w_{\alpha}(\vp,\cz),
	\qquad\qquad
	\overline{\tilde w}_{\beta,l}(\vp):=\int_0^1d\cz\:\cz^l\:\tilde w_{\beta}(\vp,\cz),
	\label{w-ab-def}\\
	&\overline{\tilde v}_{\alpha,l}(\vp):=\int_0^1d\cz\:\cz^l\:\tilde v_{\alpha}(\vp,\cz),
	\label{tilde-v=}\\
		& {X}_1(\vp,{\vp\,'}):=\frac{1}{4\pi^2\,k^2}\int_0^1\!\! d\cz_2\int_0^{\cz_2}\!\!\! d\cz_1
	\int_{\R^2} \!\!\! d^2q~\vp\cdot\vq\;\tilde v_\alpha(\cz_2,\vp-\vq)\tilde w_\alpha(\cz_1,\vq-{\vp\,'}),\\
	& {X}_2(\vp):=\frac{1}{4\pi^2}\int_0^1\!\!d\cz_2\int_0^{\cz_2}\!\!\! d\cz_1
	\int_{\R^2} \!\!\! d^2q~ \tilde w_\beta(\vp-\vq,\cz_2)\tilde w_\alpha(\vq,\cz_1),\\
	& {Y}_1(\vp,{\vp\,'}):=\frac{1}{4\pi^2\,k^2}\int_0^1\!\! d\cz_2\int_0^{\cz_2}\!\!\!d\cz_1
	\int_{\R^2}\!\!\! d^2q~\vq\cdot{\vp\,'}\,\tilde w_\alpha(\vp-\vq,\cz_2)\tilde v_\alpha(\vq-{\vp\,'},\cz_1),\\
	& {Y}_2(\vp):=\frac{1}{4\pi^2}\int_0^1\!\! d\cz_2\int_0^{\cz_2}\!\!\! d\cz_1
	\int_{\R^2} \!\!\! d^2q~ \tilde w_\alpha(\vp-\vq,\cz_2)\tilde w_\beta(\vq,\cz_1),
	\label{Y2-def}
	\end{align}
and a tilde on a function $f(\vec r,\check z)$ stands for its partial Fourier transformation with respect to $\vec r$, i.e., $\tilde f(\vec p,\check z):=\int_{\R^2} d\vec r\:e^{-i\vec p\cdot\vec r} f(\vec r,\check z)$.

As a specific example, consider the scattering of sound waves by a collection of point scatterers corresponding to $\rho(\bfr)=\rho_0$ and $\kappa(\bfr)=\sum_{i=1}^n \check\kappa_i\delta(\bfr-\bfa_j)$, where $\check\kappa_i$ are constant real or complex  parameters, and $\bfa_i\in\R^3$ are such that $0\leq \ra_{iz}\leq\ell$. Then, introducing ${\check\beta_i:=\check\kappa_i/\kappa_0}$, we have
	\begin{align}
	&\alpha(\vec r,z)=1,
	&&\beta(\vec r,z)=1+\sum_{i=1}^n\check\beta_i\delta(\vec r-\vec\ra_i)\delta(z-\ra_{iz}).	\nn
	\end{align}
These equations together with \eqref{alpha=} -- \eqref{v-alpha=} and \eqref{W-def} -- \eqref{Y2-def} imply
	\begin{align}
	&w_\alpha=v_\alpha=\tilde w_\alpha=\tilde v_\alpha=0,
	\qquad\qquad
	\overline{\tilde w}_{\alpha,l}=\overline{\tilde v}_{\alpha,l}=X_1=X_2=Y_1=Y_2=0,
	\label{pt-only-1}\\
	&W_l(\vec\rk,\vec\rk_0)=\overline{\tilde w}_{\beta,l}(\vec\rk-\vec\rk_0)=
	\frac{1}{\ell^{l+1}}\sum_{i=1}^n \check\beta_i\,\ra_{iz}^l\,e^{-i\vec\ra_i\cdot(\vec\rk-\vec\rk_0)}.
	\label{pt-only-2}
	\end{align}
Substituting {\eqref{pt-only-1} and \eqref{pt-only-2}} in \eqref{f1=} and \eqref{f2=}, we find
	\begin{align}
	\ff^{(1)}(\bn_0,\bn)&=\frac{k}{4\pi}
	W_0(\vk,\vk_0)=\frac{k}{4\pi \ell}
	\sum_{i=1}^n \check\beta_i\,e^{-i\vec\ra_i\cdot(\vec\rk-\vec\rk_0)},
	\label{f1-delta}\\
	\ff^{(2)}(\bn_0,\bn)&=\frac{ik}{4\pi}\Big[
	(\fc_0-\fc)W_1(\vk,\vk_0)
	+\frac{1}{8\pi^2}\int_{\sD_k}d^2\vq\;
	\frac{W_0(\vk,\vq)W_0(\vq,\vk_0)}{
	\sqrt{1-q^2/k^2}}\Big]
	\nn\\
    &=\frac{ik}{4\pi \ell^2}\Big[(\cos\vartheta_0-\cos\vartheta)\!
    \sum_{i=1}^n \check\beta_i\,\ra_{iz}\,e^{-i\vec\ra_i\cdot(\vec\rk-\vec\rk_0)}+
    \frac{k^2}{4\pi}\!\!\sum_{i,j=1}^{n}
    \check\beta_i\check\beta_j \,\sinc(k|\vec\ra_i-\vec\ra_j|)\,
    e^{i(\vec\ra_j\cdot\vec\rk_0-\vec\ra_i\cdot\vec\rk)}\Big],
    \label{f2-delta}
	\end{align}
where
	\[\sinc(x):=\left\{\begin{array}{ccc}
	x^{-1}\sin x&\for&x\neq 0,\\
	1&\for&x=0.\end{array}\right.\]
	
Note that, {in view of \eqref{f1-delta} and \eqref{f2-delta},} the approximation expression,
	\be
	\ff(\bn_0,\bn)\approx \ff^{(1)}(\bn_0,\bn)\,k\ell+\ff^{(2)}(\bn_0,\bn)(k\ell)^2,
	\label{f-approx}
	\ee
for the scattering amplitude is independent of $\ell$. This is actually to be expected, because for the cases where $\ra_{iz}=0$ for all $i\in\{1,2,\cdots,n\}$, $\ell$ is an arbitrary parameter with no physical significance. In Sec.~6, we obtain an exact formula for these cases and show that it agrees with the approximate expression given by \eqref{f1-delta} -- \eqref{f-approx}.

\section{Acoustic scattering by a planar array of point scatterers} 

Consider the scattering of sound waves by inhomogeneities of a compressible fluid $\mathring F$ with density $\mathring\rho$ and compressibility $\mathring\kappa$ satisfying the decay conditions at $z\to\pm\infty$ under which the bounded solutions of the corresponding Bergmann equation satisfy \eqref{asymp}. We wish to explore the consequences of the inclusion of a finite array of point scatterers in $\mathring F$ that lie on a plane orthogonal to the $z$ axis. 

{Suppose that the inclusion of the point scatterers leads to finite jumps in the fluid's density $\mr\rho$ at the point scatterers' locations (or does not affect $\mr\rho$)}, but it changes the fluid's compressibility to
	\begin{align}
	&\kappa(\vec r,z)=\mathring\kappa(\vec r,z)
	+\delta(z-\ra)\sum_{i=1}^n\check\kappa_i\delta(\vec r-\vec\ra_i),
	\label{s6-kappa}
	\end{align}
where $\check\kappa_1,\check\kappa_2,\cdots,\check\kappa_n$ are real or complex parameters, and $\bfa_i:=(\vec\ra_i,\ra)$ with $i\in\{1,2,\cdots,n\}$ are the positions of the point scatterers.\footnote{This means that $\rho$ is a continuous function of $\bfr=(\vec r,z)$ except possibly at $\bfa_i=(\vec\ra_i,\ra)$ with $i\in\{1,2,\cdots,n\}$, and that it takes finite values at these points, i.e., it has no singularities. {In other words, $\bfa_i$'s are removable discontinuities of $\rho$.}}  Since such a change in the density of the fluid results {in finitely many removable discontinuities}, it will not affect the solution of Bergmann's equation~\eqref{bergmann-eq}. Therefore, we assume, without loss of generality, that 
	\be
	\rho(\vec r,z)=\mr\rho(\vec r,z).
	\label{rho-mrrho}
	\ee 
Substituting this equation and \eqref{s6-kappa} in \eqref{a-b-def}, we find
	\begin{align}
	&\alpha(\vec r,z)=\mathring\alpha(\vec r,z),
	&&\beta(\vec r,z)=\mathring\beta(\vec r,z)+\delta(z-\ra)
	\sum_{i=1}^n\check\beta_i\delta(\vec r-\vec\ra_i),
	\label{s6-defects}
	\end{align}
where 
	\begin{align}
	&\mathring\alpha:=\frac{\mathring\rho}{\rho_0},
	&&\mathring\beta:={\frac{\mathring\kappa}{\kappa_0}},
	&&\check\beta_i:={\frac{\check\kappa_i}{\kappa_0}}.
	\label{mr-alpha-def}
	\end{align}  
Note that unlike $\mr\alpha$ and $\mr\beta$ which are dimensionless, the coefficients $\check\beta_i$ have the dimension of length$^3$.
	
\subsection{Determination of the fundamental transfer matrix}

To determine the effect of the point scatterers on the scattering of sound waves, we compute the fundamental transfer matrix of the system. As a first step in this direction, we determine the contribution  of the point scatterers to the effective Hamiltonian \eqref{bcH=}. This requires the calculation of the operators $\widehat\beta(z)$ and $\widehat\sV(z)$. 

To simplify our analysis, we confine our attention to the cases in which the point scatterers lie on the $x$-$y$ plane, i.e., set $\ra=0$. In {Appendix~A}, we derive the effect of a translation along the $z$ axis on the scattering amplitude which allows us to determine the latter for $\ra\neq 0$ from the one for $\ra=0$.

Let us introduce
	\be
	\beta'(\vec r,z):=\beta(\vec r,z)-\mr\beta(\vec r,z)=\delta(z)\sum_{i=1}^n\check\beta_i\delta(\vec r-\vec\ra_i),
	\label{beta-prime-def1}
	\ee
where we have used \eqref{s6-defects} with $\ra=0$. According to \eqref{g-id},
	\be
	\big(\widehat{\beta}'(z)\phi\big)(\vec p\,)=\delta(z)\sum_{i=1}^n\check\beta_i e^{-i\vec\ra_i\cdot\vec p}\,\widecheck\phi(\vec\ra_i),
	\label{s6-11}
	\ee
where $\widecheck\phi(\vec r)$ is the inverse Fourier transform of $\phi(\vec p)$, i.e.,
	\be
	\widecheck\phi(\vec r):=\big(\cF^{-1}\phi\big)(\vec r):=
	\frac{1}{4\pi^2}\int_{\R^2}d^2p\:e^{i\vec r\cdot\vec p}\phi(\vec p).
	\label{s6-12}
	\ee	
Since $e^{i\vec r\cdot\vec p}=2\pi\,\br\vec r\,|\vec p\,\kt$, we can use \eqref{s6-12} to express \eqref{s6-11} in the form
	\be
	\widehat{\beta}'(z)=\delta(z)\widehat{\fB}
	\label{beta-prime=}
	\ee
where
	\begin{align}
	&\widehat{\fB}:=2\pi\widehat\Delta\,\cF^{-1},
	&&\widehat\Delta:=\sum_{i=1}^n\check\beta_i|\vec\ra_i\kt\br\vec\ra_i|.
	\label{beta-prime-def}
	\end{align}

In view \eqref{eq1}, \eqref{bcH=}, \eqref{beta-prime-def1}, and \eqref{beta-prime=}, 
  		\begin{align}
	&\widehat\beta(z)=\widehat{\mr\beta}(z)+\delta(z)\widehat{\fB},\\
	&\widehat\sV(z)=\widehat{\mr\sV}(z)-k^2\delta(z)\widehat{\fB},\\
	&\widehat\bcH(z)=\widehat{\mr\bcH}(z)+\widehat{\bcH}'(z),	
	\label{s6-H=HH}
	\end{align}
where
	\begin{align}
	&\widehat{\mr\sV}(z):=\widehat{\vec p}\cdot\big[\widehat{\mr\alpha}(z)^{-1}-\widehat I\,\big]
	\widehat{\vec p}+k^2\big[\widehat I-\widehat{\mr\beta}(z)\big],
	\label{s6-sV}\\
	&\widehat{\mr\bcH}(z):=
	\frac{1}{2}e^{-iz\widehat\varpi\bsigma_3} 
	\Big\{\widehat{\mr{\sV}}(z)\widehat\varpi^{-1}\widehat\bcK
	+\widehat\varpi[\hat I-\widehat{\mr\alpha}(z)]\widehat\bcK^T\Big\}
	 e^{iz\widehat\varpi \bsigma_3},
	 \label{s6-mr-H}\\
	 &\widehat{\bcH}'(z):=-\frac{k^2}{2}\,
	\delta(z)\widehat{\fB}\widehat\varpi^{-1}\bcK.
	\label{s6-H-prime-def}
	\end{align}
	
Now, view $\widehat{\mr\bcH}(z)$ and $\widehat{\bcH}'(z)$ as ``time-dependent'' Hamiltonian operators, and let $\widehat{\mr\bcU}(z,z_0)$ and $\widehat{\bcU}'(z,z_0)$ be the corresponding evolution operators, respectively. In particular,
	\begin{align}
	&i\partial_z \widehat{\bcU}'(z,z_0)=\widehat{\bcH}'(z)
	\widehat{\bcU}'(z,z_0),
	&& \widehat{\bcU}'(z_0,z_0)=\widehat I.
	\label{ev-op-prime}
	\end{align}
Because $\bcK^2=\bzero$, $\widehat{\bcH}'(z_1)\widehat{\bcH}'(z_2)=\widehat 0$ for all $z_1,z_2\in\R$. This allows us to identify the solution of \eqref{ev-op-prime} with
	\be
	\widehat{\bcU}'(z,z_0)=\widehat I-i\int_{z_0}^z dz' \widehat{\bcH}'(z')=\widehat I+
	\frac{ik^2}{2}\,\varsigma(z,z_0)\,
	\widehat\fB\,\widehat\varpi^{-1}\widehat\bcK,
	\label{bcU-prime}
	\ee
where 
	\[\varsigma(z,z_0):=\left\{\begin{array}{cc}
	1&~\for~~z_0<0<z,\\
	-1&~\for~~z<0<z_0,\\
	0&{\rm otherwise}.
	\end{array}\right.\]

In view of \eqref{s6-H=HH} and \eqref{s6-H-prime-def}, 
	\be
	\widehat\bcH(z)=\widehat{\mr\bcH}(z)~~\for~~z\neq 0.
	\label{s6-condi1}
	\ee
Therefore, the evolution operator $\widehat\bcU(z,z_0)$ for $\widehat\bcH(z)$ satisfies
	\be
	\widehat\bcU(z,z_0)=\widehat{\mr\bcU}(z,z_0)~~\for~~z_0z>0.
	\label{s6-31}
	\ee
It is also clear from \eqref{s6-H=HH}, \eqref{bcU-prime}, and \eqref{s6-condi1} that
	\be
	\lim_{\epsilon\to 0^+}\widehat\bcU(\epsilon,-\epsilon)=
	\lim_{\epsilon\to 0^+}
	\widehat{\bcU}'(\epsilon,-\epsilon)=
	\widehat I+
	\frac{ik^2}{2}\,\widehat\fB\,\widehat\varpi^{-1}\widehat\bcK.
	\label{s6-32}
	\ee
 
We can use the composition (semi-group) property of the evolution operators together with \eqref{bcU-prime} -- \eqref{s6-32} to infer 
	\begin{align}
	\widehat\bcU(\infty,-\infty)&=
	\lim_{\epsilon\to 0^+}\big[
	\widehat\bcU(\infty,\epsilon)\,\widehat\bcU(\epsilon,-\epsilon)\,\widehat\bcU(-\epsilon,-\infty)\big]\nn\\
	&=\widehat{\mr\bcU}(\infty,0)\left[
	\widehat I+
	\frac{ik^2}{2}\,\widehat\fB\,\widehat\varpi^{-1}\widehat\bcK\right]
	\widehat{\mr\bcU}(0,-\infty)\nn\\
	&=\widehat{\mr\bcU}(\infty,-\infty)+
	\frac{ik^2}{2}\,
	\widehat{\mr\bcU}(\infty,0)\widehat\fB\,\widehat\varpi^{-1}\widehat\bcK\
	\widehat{\mr\bcU}(0,-\infty).\nn
	\end{align}
Substituting the latter equation in \eqref{M=2}, we arrive at
	\begin{align}
	\widehat\bM
	&=\widehat{\mr\bM} + \widehat\Pi_k\widehat{\mr\bfM}\, \widehat{\bfM}' \widehat\Pi_k,
	\label{M=M+MM}
	\end{align}
where 
	\begin{align}
	\widehat{\mr\bM}&:=\widehat\Pi_k\,\widehat{\mr\bcU}(\infty,-\infty)
	\widehat\Pi_k,
	\label{mr-bM}\\
	\widehat{\mr\bfM}&:=\widehat{\mr\bcU}(\infty,-\infty),
	\label{mr-bfM}\\
	\widehat\bfM'&:=\frac{ik^2}{2}\,\widehat{\mr\bcU}(0,-\infty)^{-1}
	\widehat\fB\,\widehat\varpi^{-1}\widehat\bcK\
	\widehat{\mr\bcU}(0,-\infty).
	\label{bfM-prime}
	\end{align}
Note that $\widehat{\mr\bM}$ is the fundamental transfer matrix of the fluid in the absence of the point scatterers.

In the following subsections, we investigate specific applications of Eq.~\eqref{M=M+MM}.

\subsection{Point scatterers in an effectively one-dimensional fluid}

Consider situations where $\mr\rho$ and $\mr\kappa$ -- and consequently $\mr\alpha$ and $\mr\beta$ -- are functions only of $z$. Then, in the absence of the point scatterers, the fluid has translational symmetry along the $x$ and $y$ directions, and the corresponding acoustic scattering problem is effectively one-dimensional. It is easy to see that in the case 
	\begin{align}
	&\widehat{\mr\alpha}(z)=\mr\alpha(z)\widehat I,
	&&\widehat{\mr\beta}(z)=\mr\beta(z)\widehat I.
	\end{align}
In view of \eqref{s6-sV} and \eqref{s6-mr-H}, these equations imply that 
	\begin{align}
	&\widehat{\mr\sV}(z)=\mr\sV(\widehat{\vec p},z),
	&&\widehat{\mr\bcH}(z)=\mr\bcH(\widehat{\vec p},z),
	\label{s6-eq96}
	\end{align}
where $\mr\sV$ and $\mr\bcH$ are the $2\times 2$ matrix-valued functions of $\vec p$ and $z$ that are given by
	\begin{align}
	\mr\sV({\vec p},z)&:=
	\big[{\mr\alpha}(z)^{-1}-1\,\big]\,{\vec p}\,^2
	+k^2\big[1-{\mr\beta}(z)\big]\bI,
	\label{mr-V-63}\\
	\mr\bcH({\vec p},z)&:=\frac{1}{2}e^{-iz\varpi(\vec p)\bsigma_3} 
	\Big\{{\mr{\sV}}(\vec p,z)\varpi(\vec p)^{-1} \bcK
	+\varpi(\vec p)[1-{\mr\alpha}(z)] \bcK^T\Big\}
	 e^{iz\varpi(\vec p)\bsigma_3},
	\label{mr-H-63}
	\end{align}
and $\bI$ is the $2\times 2$ identity matrix.

We can identify $\mr\bcH({\vec p},z)$ with a $2\times 2$ non-Hermitian matrix Hamiltonian which depends on ``time'' $z$ as well as a pair of real numerical parameters (coupling constants) $p_x$ and $p_y$. Let us denote the evolution operator corresponding to $\mr\bcH({\vec p},z)$ by $\mr\bcU({\vec p},z,z_0)$. This is a $2\times 2$ matrix-valued function of $\vec p:=(p_x,p_y)$, $z$, and $z_0$ that satisfies
	\begin{align}
	&i\partial_z\, \mr\bcU({\vec p},z,z_0)=\mr\bcH(\vec p,z)\,
	\mr\bcU({\vec p},z,z_0),
	&&\mr\bcU({\vec p},z_0,z_0)=\bI.
	\label{mr-U-def}
	\end{align}
Since $\bcK$ is a traceless matrix, so are $\bcK^T$ and the matrix Hamiltonian $\mr\bcH(\vec p,z)$. This in turn implies that 
	\be
	\det[\,\mr\bcU({\vec p},z,z_0)]=1.
	\label{mr-U-det}
	\ee
	
It is clear from \eqref{s6-eq96} that the evolution operator associated with the Hamiltonian operator $\widehat{\mr\bcH}(z)$ has the form
	\be
	\widehat{\mr\bcU}(z,z_0)=\mr\bcU(\widehat{\vec p},z,z_0). 
	\label{mr-bcU=}
	\ee
This equation shows that $\widehat{\mr\bcU}(z,z_0)$ commutes with $\widehat\Pi_k$. In particular, in light of \eqref{mr-bM} -- \eqref{bfM-prime}, we have
	\begin{align}
	&\widehat{\mr\bM}=\mr\bM(\widehat{\vec p}\,)\widehat\Pi_k,
	\label{mr-bM-62}\\
	&\widehat\Pi_k\widehat{\mr\bfM}=\widehat\Pi_k\widehat{\mr\bfM}\,\widehat\Pi_k=
	\widehat{\mr\bM},\\
	&\widehat\bM=\widehat{\mr\bM}\big(\widehat I+\widehat{\bsM}\big)
	\label{M=M+Ms}
	\end{align} 
where
	\begin{align}
	&\mr\bM({\vec p}\,):=\mr\bcU(\vec p,\infty,-\infty),
	\label{mr-M-def-62}
	\\
	&\widehat{\bsM}:= \frac{ik^2}{2}\,{\mr\bcU}(\widehat{\vec p},0,-\infty)^{-1}
	\widehat\Pi_k\widehat\fB\,\widehat\Pi_k\widehat\varpi^{-1}\widehat\bcK\
	{\mr\bcU}(\widehat{\vec p},0,-\infty).
	\label{s6-bsM=}
	\end{align}

We can use \eqref{mr-U-det} -- \eqref{mr-bM-62} and \eqref{mr-M-def-62} to conclude that the components $\widehat{\mr M}_{ab}$ of $\widehat{\mr\bM}$ are mutually commuting operators that commute with $\widehat\Pi_k$, and 
	\be
	\det\mr\bM(\vec p\,)=1.
	\label{mr-bM-det}
	\ee

In view of the analogy between $\mr\bM(\vec p\,)$ and the transfer matrix associated with the scattering problem defined by Bergmann's equation in one dimension, we introduce the following analogs of the reflection and transmission amplitudes in one dimension~\cite{ptep-2024b}.
	\begin{align}
	&\mr R^{l}(\vec p):=-\frac{\mr M_{21}(\vec p)}{\mr M_{22}(\vec p)}, 
	&&\mr R^{r}(\vec p):=\frac{\mr M_{12}(\vec p)}{\mr M_{22}(\vec p)}, 
	&&\mr T(\vec p):=\frac{1}{\mr M_{22}(\vec p)}.
	\label{mr-RRT}
	\end{align}
These allow us to identify \eqref{mr-bM-det} with
	\[\mr M_{11}(\vec p)=\mr T(\vec p)-\frac{\mr R^l(\vec p)\mr R^r(\vec p)}{\mr T(\vec p)}.\]
Furthermore, we employ \eqref{mr-U-det} to show that
	\be
	{\mr\bcU}({\vec p},0,-\infty)^{-1}=\left[\begin{array}{cc}
	\mr u_{22}(\vec p) & -\mr u_{12}(\vec p)\\
	-\mr u_{21}(\vec p) & \mr u_{11}(\vec p)\end{array}\right],
	\label{mr-U-inv}
	\ee
where $\mr u_{ab}(\vec p)$ stand for the entries of ${\mr\bcU}({\vec p},0,-\infty)$.

Next, we calculate the scattering amplitude $\ff(\bn_0,\bn)$ for the fluid in the presence of point scatterers. We recall that, according to \eqref{ff=} and \eqref{f-left=}, this requires the identification of the entries $\widehat M_{ab}$ of the fundamental transform matrix $\widehat\bM$ and the inversion of $\widehat M_{22}$. We have already computed $\widehat\bM$ and can use its expression, namely \eqref{M=M+Ms}, to find $\widehat M_{ab}$. As for the determination of $\widehat M_{22}^{-1}$, we observe that this is equivalent to solving the equation,
	\be
	\widehat M_{22}|\chi\kt=|\phi\kt,
	\label{int-eq}
	\ee
for $|\chi\kt\in\sF_k$, where $|\phi\kt$ is a generic element of $\sF_k$. In appendix~B, we obtain the exact solution of this equation. We express it in the form
	\be
	\widehat M_{22}^{-1}|\phi\kt:=
	|\chi\kt=\widehat{\mr T}|\phi\kt+i\pi  k^2\sum_{i,j=1}^n\check\beta_i\, \cA_{ij}^{-1}
	\Gamma_j\widehat\cZ\,\widehat\Pi_k| \vec\ra_i\kt,
	\label{M22-inv=}
	\ee
where $\cA_{ij}^{-1}$ are the entries of the inverse of the $n\times n$ matrix $\bcA:=[\cA_{ij}]$, 
	\begin{align}
	&\cA_{ij}:=\delta_{ij}-i\pi k^2\check\beta_j\br \vec\ra_i|\cF^{-1}
	\widehat\varpi^{-1}\widehat\cX\widehat\cZ\,\widehat
	\Pi_k| \vec\ra_j\kt=\delta_{ij}-i \alpha_{ij}\check\beta_j,
	\label{cA-ij=}\\[6pt]
	&\alpha_{ij}:=\frac{k^2}{8\pi^2}\int_{\sD_k}d^2 p\: 
	\frac{e^{i( \vec\ra_i- \vec\ra_j)\cdot\vec p}\cX(\vec p\,)\cZ(\vec p\,)}{\varpi(\vec p)},
	\label{alpha-ij}\\
	&\Gamma_j:=\br \vec\ra_j|\cF^{-1}\widehat\cX\,\widehat{\mr T}\widehat\varpi^{-1}|\phi\kt =\frac{1}{4\pi^2}\int_{\sD_k}d^2p\: 
	\frac{e^{i \vec\ra_j\cdot\vec p}\cX(\vec p\,)\mr T(\vec p\,)\phi(\vec p\,)}{\varpi(\vec p\,)},
	\label{gamma-j}\\[3pt]
	&\widehat\cX:=\cX(\widehat{\vec p}\,),
	\qquad\qquad\qquad\qquad\qquad\cX(\vec p\,):=\mr u_{12}({\vec p}\,)+\mr u_{22}({\vec p}\,),
	\label{cX-def}\\
	&\widehat\cZ:=\cZ(\widehat{\vec p}\,), 
	\qquad\qquad\qquad\qquad\qquad
	\cZ(\vec p\,):={\mr R}^l(\vec p\,)\cX(\vec p\,)+\cY(\vec p\,),
	\label{cZ-def}\\
	&\cY(\vec p\,):=\mr u_{11}({\vec p}\,)+\mr u_{21}({\vec p}\,),
	\qquad\qquad 
	\widehat{\mr T}:= {\mr T}(\widehat{\vec p}\,).
	\label{hat-mr-T}
	\end{align}
It is also worth mentioning that
	\begin{align}
	&\widehat\cZ\,\widehat\Pi_k| \vec\ra_i\kt=\frac{1}{2\pi}\int_{\sD_k}d^2p\: e^{-i \vec\ra_i\cdot\vec p}
	\cZ(\vec p\,)|\vec p\kt.
	\label{b-j}
	\end{align}
Moreover, given a function $f:\sD_k\to\C$, we have
	{\be
	\int_{\sD_k}d^2p\:\frac{f(\vec p\,)}{\varpi(\vec p\,)}=
	k\int_0^{\pi/2}\!\!d\theta\int_0^{2\pi}\!\!d\varphi\: \sin\theta\,f(k\sin\theta\, \cos\varphi,k\sin\theta\, \sin\varphi),
	\label{id-100}
	\ee
where $f(k\sin\theta\,\cos\varphi,k\sin\theta\,\sin\varphi)$ stands for $f(p_x,p_y)$ with $p_x=k\sin\theta \cos\varphi$ and $p_y=k\sin\theta\, \sin\varphi$.}

Eq.~\eqref{M22-inv=} gives the solution of \eqref{int-eq} provided that $\mr R^l$ and $\mr T$ do not have singularities, i.e., $\mr M_{22}(\vec p\,)\neq 0$ for all $\vec p\in\sD_k$, and $\bcA$ is an invertible matrix. For generic values of $\check\beta_i$ and $k$, the latter condition holds. However, in general, there may exist values of $\check\beta_i$'s and $k$ for which $\det\bcA=0$ and $\bcA$ becomes singular. These mark the spectral singularities of the system \cite{prl-2009} at which the scattering amplitude diverges and the system begins emitting coherent acoustic waves, a phenomenon called SASER (sound amplification by stimulated emission of radiation) \cite{kend-2006}. 

Having determined $\widehat M_{22}^{-1}$, we can use \eqref{ff=} and \eqref{f-left=} to derive an explicit formula for the scattering amplitude $\ff(\bn_0,\bn)$. To do this, first we express $\ff(\bn_0,\bn)$ in the form
	\be
	\ff(\bn_0,\bn)= \left\{\begin{array}{ccc}
	\ff_{+,+}&\for&\rn_{0z}>0~{\rm and}~\rn_z>0,\\
	\ff_{+,-}&\for&\rn_{0z}>0~{\rm and}~\rn_z<0,\\
	\ff_{-,+}&\for&\rn_{0z}<0~{\rm and}~\rn_z>0,\\
	\ff_{-,-}&\for&\rn_{0z}<0~{\rm and}~\rn_z<0,
	\end{array}\right.
	\label{ff=defect}
	\ee
where, in view of \eqref{ff=} and \eqref{f-left=},
	\begin{align}
	&\ff_{+,+}:=-2\pi i\varpi(\vec\rk)\big[\br-\vec\rk_0|\widehat M_{22}^{-1}
	|\!-\!\vec\rk\kt-\delta(\vec\rk-\vec\rk_0)\big],
	\label{fpp=}\\
	&\ff_{+,-}:=2\pi i\varpi(\vec\rk_0)\,\br\vec\rk|\widehat M_{22}^{-1}
	\widehat M_{21}|\vec\rk_0\kt,
	\label{fpm=}\\
	&\ff_{-,+}:=-2\pi i\varpi(\vec\rk_0)\,\br\vec\rk|\widehat M_{12}
	\widehat M_{22}^{-1}|\vec\rk_0\kt,
	\label{fmp=}\\
	&\ff_{-,-}:=-2\pi i\varpi(\vec\rk_0)\big[\br\vec\rk|\widehat M_{22}^{-1}
	|\vec\rk_0\kt-\delta(\vec\rk-\vec\rk_0)\big].
	\label{fmm=}
	\end{align}
We present the details of the calculations of the right-hand sides of these relations in Appendix~C. Here we give their final result {which turn out to be remarkably simple}:	 
	{\begin{align}
	\ff_{+,+}&=\mr\ff_{+,+}+Q(\vec\rk,\vec\rk_0)\,\mr T(\vec\rk) \cX(\vec\rk)\cZ(\vec\rk_0),
	\label{fpp-0}\\[3pt]
	\ff_{+,-}
	&=\mr\ff_{+,-}+Q(\vec\rk,\vec\rk_0)\,\cZ(\vec\rk) \cZ(\vec\rk_0),
	\label{fpm-0}\\
	\ff_{-,+}
	&=\mr\ff_{-,+}+Q(\vec\rk,\vec\rk_0)\,
	\mr T(\vec\rk)\cX(\vec\rk)\mr T(\vec\rk_0)\cX(\vec\rk_0),
	\label{fmp-0}\\[3pt]
	\ff_{-,-}&=\mr\ff_{+,+}+Q(\vec\rk,\vec\rk_0)\,
	\cZ(\vec\rk) \mr T(\vec\rk_0) \cX(\vec\rk_0) ,
	\label{fmm-0}
	\end{align}}
where
	\begin{align}
	&\mr\ff_{+,+}:=-2\pi i\, \varpi(\vec\rk_0)\delta(\vec\rk-\vec\rk_0)[\mr T(\vec\rk_0)-1],
	&&\mr\ff_{+,-}:=- 2\pi i\,\varpi(\vec\rk_0)\delta(\vec\rk-\vec\rk_0) \mr R^l(\vec\rk_0),
	\label{mr-fp}\\[3pt]
	&\mr\ff_{-,+}:=-2\pi i\,\varpi(\vec\rk_0)\delta(\vec\rk-\vec\rk_0) \mr R^r(\vec\rk_0),
	&&Q(\vec\rk,\vec\rk_0):=\frac{k^2}{4\pi}\sum_{i,j=1}^n\check\beta_i\,\cA_{ij}^{-1}e^{-i(\vec\ra_i\cdot\vec\rk-\vec\ra_j\cdot\vec\rk_0)}.
	\label{mr-fm}
	\end{align}
	
When the point scatterers are absent, $\check\beta_i=0$ and \eqref{mr-fm} imply $Q(\vec\rk,\vec\rk_0)=0$. Inserting this equation in \eqref{fpp-0} -- \eqref{fmm-0} and using \eqref{ff=defect}, we find 
 	\be
	\ff(\bn_0,\bn)=\mr\ff(\bn_0,\bn):= \left\{\begin{array}{ccc}
	\mr\ff_{+,+}&\for&\rn_{0z}\rn_z>0,\\
	\mr\ff_{+,-}&\for&\rn_{0z}>0~{\rm and}~\rn_z<0,\\
	\mr\ff_{-,+}&\for&\rn_{0z}<0~{\rm and}~\rn_z>0.
	\end{array}\right.
	\label{mr-ff}
	\ee
Therefore, $\mr\ff(\bn_0,\bn)$ is the scattering amplitude of the fluid in the absence of point scatterers.

Another useful observation is that $\mr R^{l/r}$, $\mr T$, $\cX$, and $\cZ$ which enter the expression for the scattering amplitude are functions of $\varpi(\vec p)$. This follows from the fact that the fluid $\mr F$ has {axial symmetry about the $z$ axis}. If we respectively parameterize the unit vectors $\bn_0$ and $\bn$ in terms of the corresponding polar and azimuthal angular coordinates $(\vartheta_0,\varphi_0)$ and $(\vartheta,\varphi)$, so that
	\begin{align}
	&\bn_0=(\sin\vartheta_0 \cos\varphi_0,\sin\vartheta_0\sin\varphi_0,\cos\vartheta_0),
	&&\bn=(\sin\vartheta \cos\varphi,\sin\vartheta\sin\varphi,\cos\vartheta),
	\label{angles}
	\end{align}
then $\varpi(\vec\rk_0)=k|\cos\vartheta_0|$ and $\varpi(\vec\rk)=k|\cos\vartheta|$. This in turn implies that all the terms appearing on the right-hand sides of \eqref{fpp-0} -- \eqref{fmm-0} are functions only of $\vartheta_0$ and $\vartheta$ except for $Q(\vec\rk,\vec\rk_0)$. In other words, the $\varphi_0$- and $\varphi$- dependence of the scattering amplitude is determined by $Q(\vec\rk,\vec\rk_0)$.

Since $\cX$ and $\cZ$ are functions of $\varpi(\vec p)$, and $\varpi$ is a function of $|\vec p|$, the same applies to $\cX$ and $\cZ$. If we express the integral in \eqref{alpha-ij} in the polar coordinate in the $p_x$-$p_y$ plane, we can use this property of $\cX$ and $\cZ$ to evaluate the resulting angular integral;	\begin{align}
	\alpha_{ij}&=\frac{k^2}{8\pi^2}\int_0^k dp \int_{0}^{2\pi} d\gamma\:
	\frac{p\:e^{i|\vec\ra_i-\vec\ra_j|p\cos\gamma}\cX(\vec p\,)\cZ(\vec p\,)}{\varpi(\vec p\,)}\nn\\
	&=\frac{k^2}{4\pi}\int_0^k dp\:\frac{p\, J_0\big(|\vec\ra_i-\vec\ra_j|p\big)\sX(\sqrt{k^2-p^2})\sZ(\sqrt{k^2-p^2})}{\sqrt{k^2-p^2}}\nn\\
	&={\frac{k^3}{4\pi}\int_{0}^{\pi/2}d\theta\:\sin\theta\,J_0\big(|\vec\ra_i-\vec\ra_j|k\sin\theta\big)\sX(k\cos\theta)\sZ(k\cos\theta),}
	\label{alpha-ij-sym}
	\end{align}
where  $J_0$ is the zeroth-order Bessel function of the first kind, we have made use of the identity, $\int_0^{2\pi} d\gamma\,e^{i x\cos\gamma}=2\pi J_0(x)$ which holds for all $x\in\R$, and introduced the functions {$\sX:(0,k]\to\C$ and $\sZ:(0,k]\to\C$ given by $\sX(w):=\cX(\sqrt{k^2-w^2},0)$ and $\sZ(w):=\cZ(\sqrt{k^2-w^2},0)$, which for all $\vec p\in\sD_k$ and $p:=|\vec p\,|$ satisfy
	\begin{align}
	&\sX(\sqrt{k^2-p^2})=\sX(\varpi(\vec p\,))=\cX(\vec p),
	&&\sZ(\sqrt{k^2-p^2})=\sZ(\varpi(\vec p\,))=\cZ(\vec p).
	\label{sX-sZ}
	\end{align}
}%

\subsection{Point scatterers in a homogeneous fluid}

If the carrier fluid $\mr F$ is homogeneous, $\mr\rho=\rho_0$, $\mr\kappa={\kappa_0}$, and \eqref{mr-alpha-def}, \eqref{mr-V-63} -- \eqref{mr-U-def}, and 
\eqref{mr-M-def-62} give
	\begin{align}
	&\mr\alpha(z)=\mr\beta(z)=1, &&\mr\bcH(\vec p,z)=\mr\sV(\vec p,z)=\bzero, &&\mr\bcU(\vec p,z,z_0)=\mr\bM=\bI.
	\end{align} 
Substituting these equations in \eqref{mr-RRT}, \eqref{alpha-ij}, \eqref{cX-def} -- \eqref{hat-mr-T}
and making use of \eqref{id-100}, we obtain
	\begin{align}
	&\mr R^{l/r}=0,
	\qquad\quad
	\mr T=\cX=\cY=\cZ=1,
	\label{6-3-eq1a}\\
	&\alpha_{ij}=\frac{k^3}{4\pi}\,\sinc(k|\vec\ra_i-\vec\ra_j|).
	\label{6-3-eq1b}
	\end{align}
Using \eqref{6-3-eq1a} in \eqref{fpp-0} -- \eqref{mr-fm}, we can express \eqref{ff=defect} in the form
	\be
	\ff(\bn_0,\bn)=Q(\vec\rk,\vec\rk_0).
	\label{Q-meaning}
	\ee

Having obtained an explicit formula for the scattering amplitude, we can easily evaluate its low-frequency behavior.  To do this, first we use \eqref{cA-ij=} and \eqref{6-3-eq1b} to show that
	\be
	\cA_{ij}=\delta_{ij}-\frac{ik^3\,\check\beta_j}{4\pi}\,\sinc(k|\vec\ra_i-\vec\ra_j|).
	\label{6-3-eq1c}
	\ee
This implies that
	\be
	\cA_{ij}^{-1}=\delta_{ij}+\frac{ik^3\,\check\beta_j}{4\pi}\,\sinc(k|\vec\ra_i-\vec\ra_j|)+\cO(k^5),
	\label{cA-inv}
	\ee
where $\cO(k^m)$ stands for a power series whose leading-order term is of the order $k^m$ or higher. Substituting \eqref{cA-inv} in the second equation in \eqref{mr-fm} and using \eqref{Q-meaning}, we have
	\be
	\ff(\bn_0,\bn)=\frac{k^2}{4\pi}\left[\sum_{i=1}^n \check\beta_i
	e^{-i\vec\ra_i\cdot(\vec\rk-\vec\rk_0)}+
	\frac{ik^3}{4\pi}\sum_{i,j=1}^n\check\beta_i\check\beta_j \,\sinc(k|\vec\ra_i-\vec\ra_j|)\,e^{-i(\vec\ra_i\cdot\vec\rk-\vec\ra_j\cdot\vec\rk_0)}\right]+\cO(k^7).
	\ee
This equation is in perfect agreement with the outcome of the low-frequency approximation given by \eqref{f1-delta} -- \eqref{f-approx} with $\ra_{iz}=0$.

\subsection{Point scatterers contained in a homogeneous acoustic slab}

\begin{figure}
    \begin{center}
        \includegraphics[scale=.2]{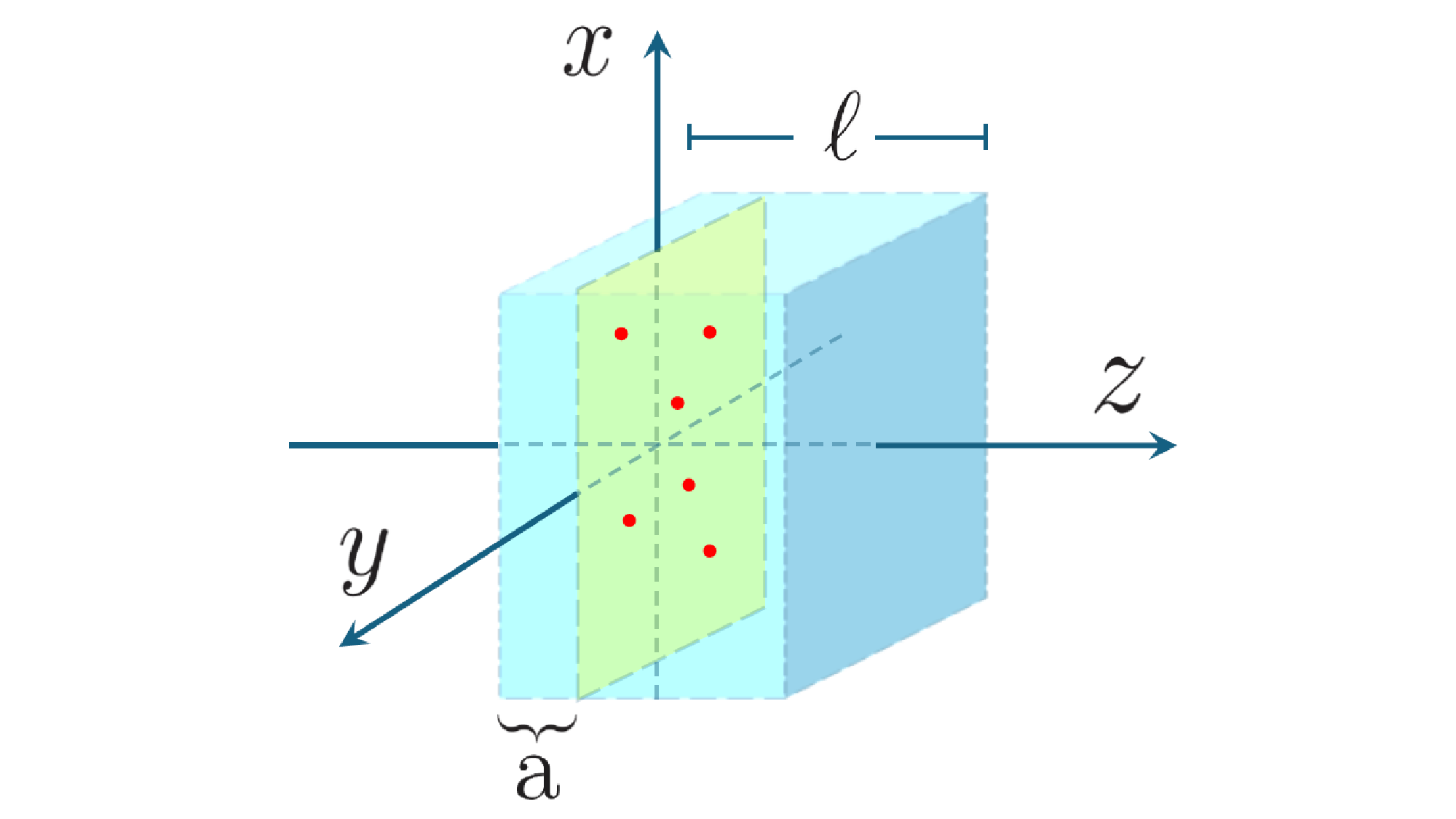}
        \caption{Schematic view of an acoustic slab of thickness $\ell$ containing delta-function point scatterers lying on the $x$-$y$ plane (marked in green). The red dotes on this plane stand for the point scatterers. The blue region corresponding to $-\ra\leq z\leq\ell-\ra$ represents the slab.}
        \label{fig2}
    \end{center}
    \end{figure}

Consider the cases where $\mr F$ consists of a homogeneous acoustic slab $\mr\cS$ of thickness $\ell$ that contains a planar array of delta-function point scatterers lying on the $x$-$y$ plane. Let $\rho_s$, $\kappa_s$, and $\fn_s$ respectively denote the density, compressibility, and refractive index of the slab in the absence of the point scatterers, and $\ra\in[0,\ell]$ be such that the planes given by $z=-\ra$ and $z=\ell-\ra$ coincide with the faces of the slab, as shown in Fig.~\ref{fig2}. 
Then,
	\begin{align}
	&\mr\rho(\vec r,z):=\left\{\begin{array}{cc }
	\rho_s &\for~-\ra\leq z \leq\ell-\ra,\\
	\rho_0&\mbox{otherwise},\end{array}\right.
	&&\mr\kappa(\vec r,z):=\left\{\begin{array}{cc }
	\kappa_s &\for~ -\ra\leq z \leq\ell-\ra,\\
	{\kappa_0}&\mbox{otherwise}.\end{array}\right.
	\label{s64-z1}\\
	&\alpha(\vec r,z)=\mr\alpha(\vec r,z)=\left\{\begin{array}{cc }
	\alpha_s &\for~ -ra\leq z \leq\ell-\ra,\\
	1&\mbox{otherwise},\end{array}\right.
	&&\mr\beta(\vec r,z)=\left\{\begin{array}{cc }
	\beta_s &\for~ -\ra\leq z \leq\ell-\ra,\\
	1&\mbox{otherwise},\end{array}\right.
	\label{s64-z2}\\
	&\beta(\vec r,z)=\mr\beta(\vec r,z)+\delta(z)\sum_{i-1}^n\check\beta_i\delta(\vec r-\vec\ra_i),
	\label{s64-z3}
	\end{align}
where 
	\begin{align}
	&\alpha_s:=\frac{\rho_s}{\rho_0},
	&&\beta_s:={\frac{\kappa_s}{\kappa_0}}=\frac{\fn_s^2}{\alpha_s}.
	\end{align}
 
 To compute the scattering amplitude for sound waves incident on the slab $\mr\cS$, we need to compute the matrices $\mr\bM(\vec p)$ and $\mr\bcU(\vec p,0,-\infty)$ whose entries enter the calculation of $\mr R^{l/r}$, $\mr T$, $\cX$, and $\cY$. This requires the calculation of the evolution operator $\mr\bcU(\vec p,z,z_0)$ associated with the matrix Hamiltonian $\mr\bcH(\vec p,z)$. We give the details of this calculation in Appendix D. The following is the resulting formulas for the entries $\mr\cU_{ab}(\vec p,z,z_0)$ of $\mr\bcU(\vec p,z,z_0)$.
	\begin{align}
	\mr\cU_{11}(\vec p,z,z_0)&:= e^{-i\varpi(\vec p)(z-z_0)}
	\big[\fc(\vec p,z-z_0)+i\fm_+(\vec p\,)	\fs(\vec p,z-z_0)\big],
	\label{cU11=}\\
	\mr\cU_{12}(\vec p,z,z_0)&:=i e^{-i\varpi(\vec p)(z+z_0)}\fm_-(\vec p\,)\fs(\vec p,z-z_0),
	\label{cU12=}\\
	\mr\cU_{21}(\vec p,z,z_0)&:=-i e^{i\varpi(\vec p)(z+z_0)}\fm_-(\vec p\,)\fs(\vec p,z-z_0),
	\label{cU21=}\\
	\mr\cU_{22}(\vec p,z,z_0)&:=e^{i\varpi(\vec p)(z-z_0)}
	\big[\fc(\vec p,z-z_0)-i\fm_+(\vec p\,)	\fs(\vec p,z-z_0)\big],
	\label{cU22=}
	\end{align}
where	
	\begin{align}
	&\fc(\vec p,z):=\cos[\tilde\fn(\vec p\,)\varpi(\vec p\,)z],
	&&\fs(\vec p,z):=\tilde\fn(\vec p\,)^{-1}\sin[\tilde\fn(\vec p\,)\varpi(\vec p\,)z],
	\label{fc-fs-def}\\
	&\tilde{\fn}(\vec p\,):=\pm\sqrt{\frac{k^2(\alpha_s\beta_s-1)}{\varpi(\vec p\,)^{2}}+1},
	&&\fm_{\pm}(\vec p\,):= \frac{\tilde{\fn}(\vec p\,)^2\pm\alpha_s^2}{2\alpha_s},
	\label{m-tn-def}
	\end{align}
where the $\pm$ in the formula for $\tilde\fn$ is the sign of the real part of the refractive index $\fn_s$ of the slab.{\footnote{{Note that $\fn_s^2=\alpha_s\beta_s$.}}} It is positive unless the slab is made of a negative-index metamaterial \cite{acoustic-meta1,acoustic-meta2}, {in which case the real parts of $\alpha$ and $\beta$ are also negative.}
	
Next, we recall that according to \eqref{s64-z2}, $\alpha(\vec r,a)=\beta(\vec r,z)=1$ for $z<-\ra$ and $z>\ell-\ra$. This observation together with \eqref{mr-V-63} and \eqref{mr-H-63} imply that $\mr\bcH(\vec p,z)=\bzero$ for these values of $z$. As a results, $\mr\bcU(\vec p,z,z_0)=\bI$ for $z_0\leq z\leq-\ra$ and $z\geq z_0\geq\ell-\ra$, {and \eqref{mr-U-def}, \eqref{mr-M-def-62}, and \eqref{mr-H-6s-a} imply}
 	\begin{align}
	&\mr\bM({\vec p}\,):=\mr\bcU(\vec p,\ell-\ra,-\ra),
	&&\mr\bcU(\vec p,0,-\infty)=\mr\bcU(\vec p,0,-\ra).
	\label{mrU-mrM=}
	\end{align}
Equations \eqref{mr-RRT}, \eqref{cU11=} -- \eqref{cU22=},  and \eqref{mrU-mrM=} allow us to derive the following formulas for $\mr R^{l/r}(\vec p)$, $\mr T(\vec p)$, and the entries $\mr u_{ab}(\vec p)$ of $\mr\bcU(\vec p,0,-\infty)$.
	\begin{align}
	&\mr R^l(\vec p\,)=
	\frac{i \fm_-(\vec p\,)\,e^{-2i\,\ra\,\varpi(\vec p\,)}}{\ft(\vec p,\ell)},
	&&\mr R^r(\vec p\,)=
	\frac{i \fm_-(\vec p\,)\,e^{-2i(\ell-\ra)\varpi(\vec p\,)}}{\ft(\vec p,\ell)},
	\label{mr-RR=s}\\
	&\mr T(\vec p\,)=
	\frac{e^{-i\varpi(\vec p)\ell}}{\fs(\vec p,\ell)\,\ft(\vec p,\ell)},
	&&\ft(\vec p,z):=\frac{\fc(\vec p,z)}{\fs(\vec p,z)}-i\fm_+(\vec p\,),
	\label{mr-T=s}
	\\[3pt]
	&\mr u_{11}(\vec p\,)= e^{-i\,\ra\,\varpi(\vec p)}\big[\fc(\vec p,\ra)+i\fm_+(\vec p\,)\fs(\vec p,\ra)\big],
	&&\mr u_{12}(\vec p\,)=i e^{i\,\ra\,\varpi(\vec p)}\fm_-(\vec p\,)\fs(\vec p,\ra),
	\label{u1=}\\[3pt]
	&\mr u_{21}(\vec p\,)=-i e^{-i\,\ra\,\varpi(\vec p)}\fm_-(\vec p\,)\fs(\vec p,\ra),
	&&\mr u_{22}(\vec p\,)=
	e^{i\,\ra\,\varpi(\vec p)}\big[\fc(\vec p,\ra)-i\fm_+(\vec p\,)\fs(\vec p,\ra)\big].	
	\label{u22=}
	\end{align}
Substituting \eqref{mr-RR=s}, \eqref{u1=}, and \eqref{u22=} in \eqref{cX-def} -- \eqref{hat-mr-T} and making use of \eqref{mr-T=s}, we find	
	\begin{align}
	&\cX(\vec p\,) =e^{i\ra\,\varpi(\vec p\,)}\big[\fc(\vec p,\ra)-i\alpha_s\,\fs(\vec p,\ra)\big],
	&&\mr T(\vec p\,) \cX(\vec p\,) = e^{-i(\ell-\ra)\varpi(\vec p\,)}\!
	\left[\frac{\fc(\vec p,\ra)-i\alpha_s\fs(\vec p,\ra)}{
	\fc(\vec p,\ell)-i\fm_+(\vec p)\,\fs(\vec p,\ell)}\right]\!,
	\label{X1-XT=}\\
	&\cY(\vec p\,)  =e^{-i\ra\,\varpi(\vec p\,)}\big[\fc(\vec p,\ra)+i\alpha_s\,\fs(\vec p,\ra)\big],
	&&\cZ(\vec p\,)  = \frac{e^{-i\ra\,\varpi(\vec p\,)}
	\big[\fc(\vec p,\ell-\ra)-i\alpha_s\,\fs(\vec p,\ell-\ra)\big]}{\fc(\vec p,\ell)-i\fm_+(\vec p)\,\fs(\vec p,\ell)}.
	\label{Y1-Z1=}
	\end{align}
	
Having computed $\cX$ and $\cZ$, we can use \eqref{alpha-ij} and \eqref{mr-fm} to evaluate the values of the coefficients $\alpha_{ij}$, find and invert the matrix $\bcA$, and determine the function $Q(\vec\rk,\vec\rk_0)$. This completes the calculation of the terms appearing on the right-hand sides of \eqref{fpp-0} -- \eqref{mr-fm} which together with \eqref{ff=defect} give the scattering amplitude for the system.

{Equations \eqref{X1-XT=} and \eqref{Y1-Z1=} simplify considerably for $\ra=0$. This corresponds to cases where the slab is placed between the planes {defined by} $z=0$ and $z=\ell$, and the point scatterers reside on the former plane. In this case, \eqref{mr-RR=s}, \eqref{X1-XT=}, and \eqref{Y1-Z1=} give  
	\begin{align}
	&\cX(\vec p\,)=1,
	&&\cZ(\vec p\,)=\mr R^l(\vec p)+1=\mr R(\vec p)+1,
	\label{a-zero}
	\end{align}
where
	\be
	\mr R(\vec p):=\frac{i\fm_-(\vec p\,)}{\ft(\vec p,\ell)}.
	\label{R-a-zero}
	\ee	
Similarly, for $\ra=\ell$, we have 
	\begin{align}
	&\cX(\vec p\,)=\frac{\mr R^r(\vec p\,)+1}{\mr T(\vec p\,)}=\frac{\mr R(\vec p\,)+1}{\mr T(\vec p\,)},
	&&\cZ(\vec p\,)= \mr T(\vec p\,).
	\label{a-ell}
	\end{align}}
	
{Substituting \eqref{a-zero} and \eqref{a-ell} in \eqref{alpha-ij-sym}, we find that, for {both} $\ra=0$ and $\ra=\ell$,
	\begin{align}
	&\alpha_{11}=\alpha_{22}=\frac{k^3}{4\pi}(	\overline{R}_1+1),
	&&\alpha_{12}=\alpha_{21}=\frac{k^3}{4\pi}\,
	\big[\,\overline{R}_2+\sinc(dk)\big] 
	\label{alpha-ij-a-zero}
	\end{align}
where
	\begin{align}
	&\overline{R}_1:=\int_0^{\pi/2}d\theta\,\sin\theta\, \mr\sR (\theta),
	\qquad\quad
	\overline{R}_2:=\int_0^{\pi/2}d\theta\,\sin\theta\, J_0(kd\sin\theta)\,\mr\sR (\theta),
	\label{overline-Rs}\\
	&\mr\sR(\theta):=\mr R(\vec p)=\mr R(k\cos\theta,0)\,~~\for~~|\vec p\,|=k\sin\theta.
	\label{sR-a-zero}
	\end{align}}


\subsection{Acoustic slab with two point scatterers on one of its faces}

\begin{figure}
   { \begin{center}
        \includegraphics[scale=.3]{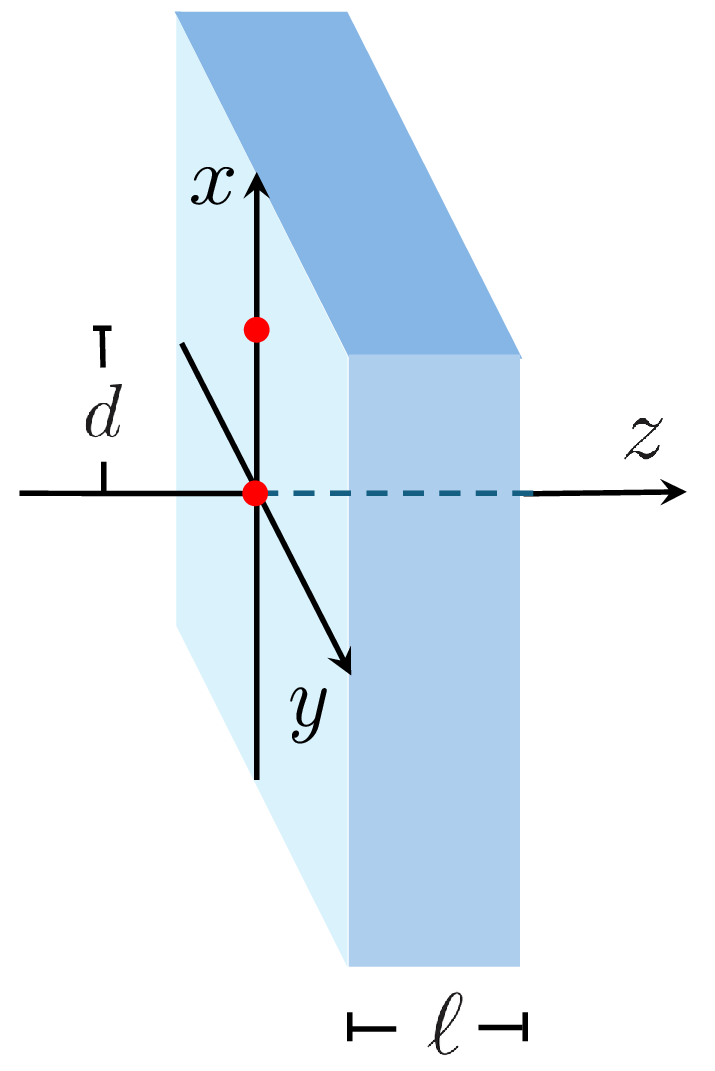}
        \caption{Schematic view of a homogeneous acoustic slab containing a pair of delta-function point scatterers (shown as red balls) on one of its faces.}
        \label{fig3}
    \end{center}}
    \end{figure}

Suppose that there is a pair of point scatterers at $\bfa_1=(0,0,0)$ and $\bfa_2=(d,0,0)$ for some real parameter $d$, {as shown in Fig.~\ref{fig3}.}
{Then, 
	\begin{align}
	&\ra=0, &&\vec\ra_1=(0,0), &&\vec\ra_2=(d,0),
	\label{ras=}
	\end{align}
\eqref{a-zero} holds, and \eqref{ff=defect} and \eqref{fpp-0} -- \eqref{fmm-0} give	
	\begin{align}
	\ff(\bn_0,\bn)&=\mr\ff(\bn_0,\bn)+\left\{\begin{array}{ccc}
	Q(\vec\rk,\vec\rk_0)[\mr R(\vec\rk_0)+1]
	\mr T(\vec\rk)&\for&\rn_{0z}>0~{\rm and}~\rn_z>0,\\
	Q(\vec\rk,\vec\rk_0)[\mr R(\vec\rk)+1][\mr R(\vec\rk_0)+1]&\for&\rn_{0z}>0~{\rm and}~\rn_z<0,\\
	Q(\vec\rk,\vec\rk_0)\mr T(\vec\rk)\mr T(\vec\rk_0)&\for&\rn_{0z}<0~{\rm and}~\rn_z>0,\\
	Q(\vec\rk,\vec\rk_0)[\mr R(\vec\rk)+1]\mr T(\vec\rk_0) &\for&\rn_{0z}<0~{\rm and}~\rn_z<0.
	\end{array}\right.
	\label{f-a=zero}
	\end{align}}%
To obtain more explicit expressions for $\mr R$ and $\mr T$, we use \eqref{m-tn-def} to show that, for $|\vec p\,|=k\cos\theta$,
	\begin{align}
	&\tilde\fn(\vec p\,)=\pm|\sec\theta|\,\nu(\theta)
	&&\fm_\pm(\vec p\,)=\mu_\pm(\theta),
	\label{tn-m=expl}
	\end{align}
where
	\begin{align}
	&\nu(\theta):=\sqrt{\fn_s^2-\sin^2\theta},
	&\mu_\pm(\theta):=\frac{(\fn_s^2-1)\tan^2\theta+\fn_s^2\pm\alpha_s^2}{2\alpha_s}.
	\end{align}
Since in this case $\varpi(\vec p)=k|\cos\theta|$, we have 
$\ell\varpi(\vec p)\tilde\fn(\vec p\,)=\pm k\ell\,\nu(\theta)$. This equation together with \eqref{mr-T=s}, \eqref{R-a-zero}, \eqref{sR-a-zero}, and \eqref{tn-m=expl} imply
	\begin{align}
	&\mr T(\vec p)=\mr\sT(\theta):=\frac{i\nu(\theta)\,e^{-ik\ell |\cos(\theta)|}}{
	\nu(\theta)\cos[k\ell\,\nu(\theta)]	-i|\cos\theta|\,\mu_+(\theta)\sin[k\ell\,\nu(\theta)]},
	\label{sT-explicit}\\
	&\mr R(\vec p\,)=\mr\sR(\theta)= \frac{i |\cos\theta|\,\mu_-(\theta)}{
	\nu(\theta)\cot[k\ell\,\nu(\theta)]-i|\cos\theta|\,\mu_+(\theta)}.
	\label{sR-explicit}
	\end{align}
Notice that 
	\begin{align}
	&\mr T(\vec\rk\,)=\mr\sT(\vartheta),
	&&\mr T(\vec\rk_0\,)=\mr\sT(\vartheta_0),
	&&\mr R(\vec\rk\,)=\mr\sR(\vartheta),
	&&\mr R(\vec\rk_0\,)=\mr\sR(\vartheta_0),
	\nn
	\end{align}
where $\vartheta$ and $\vartheta_0$ are respectively the polar angular coordinates of the unit vectors $\bn$ and $\bn_0$. 

Comparing Eqs.~\eqref{sT-explicit} and \eqref{sR-explicit} with Eqs.~(30) and (31) of Ref.~\cite{jpa-2025}, we can identify $\mr\sR(\theta)$ and $\mr\sT(\theta)$ with the left reflection and transmission amplitude of the homogeneous for an incident wave with incidence angle $\theta$, while the right reflection amplitude of the slab coincides with
	\be
	\mr R^r(\vec p)=\mr\sR(\theta)\, e^{2ik\ell |\cos\theta|}.
	\nn
	\ee
	
{Next, we recall that \eqref{ras=} implies \eqref{alpha-ij-a-zero}. Substituting this relation in \eqref{cA-ij=}, we find
	\begin{align}
	&\bcA=\left[\begin{array}{cc}
	1-i\fz_1(\overline{R}_1+1) & -i\fz_2[\overline{R}_2+\sinc(dk)] \\
	-i\fz_1[\overline{R}_2+\sinc(dk)] & 1-i\fz_2(1+\overline{R}_1) \end{array}\right],\\
	&\bcA^{-1}:=\frac{1}{D}
	\left[\begin{array}{cc}
	1-i\fz_2(\overline{R}_1+1) & i\fz_2[\overline{R}_2+\sinc(dk)] \\
	i\fz_1[\overline{R}_2+\sinc(dk)] & 1-i\fz_1(\overline{R}_1+1) \end{array}\right],
	\label{A-A-inv}
	\end{align}
where we have introduced 
	\be
	\fz_i:=\frac{k^3\check\beta_i}{4\pi},
	\label{fz-i=}
	\ee
and
	\begin{align}
	&D:=\det(\bcA)=1-i(\fz_1+\fz_2)(\overline{R}_1+1)+\fz_1\fz_2\big\{
	[\overline{R}_2+\sinc(dk)]^2-(\overline{R}_1+1)^2\big\}.
	\label{fz-def}
	\end{align}
	}
	
Having computed $\bcA^{-1}$, we can determine its entries. Substituting these in \eqref{mr-fm} and using \eqref{ras=}, we obtain 
	\begin{align}
	Q(\vec\rk,\vec\rk_0)=&\frac{1}{kD}\Big\{\fz_1+\fz_2\, e^{-id(\rk_x-\rk_{0x})}+
	i\fz_1\fz_2\Big[[\overline{R}_2+\sinc(dk)]( e^{id\rk_{0x}}+e^{-id\rk_x})\nn\\&
	\hspace{5.8cm}-(\overline{R}_1+1)(1+e^{-id(\rk_x-\rk_{0x}})\Big]\Big\},
	\label{Q-a-zero}
	\end{align}
where $\rk_x$ and $\rk_{0x}$ are respectively the $x$ components of $\vec\rk$ and $\vec\rk_0$. Since $\vec\rk=k\vec\rn$ and $\vec\rk_0=k\vec\rn_0$, we can use \eqref{angles} to express $\rk_x$ and $\rk_{0x}$ in the form
	{\begin{align}
	&\rk_x=k\,\rn_x=k\sin\vartheta\cos\varphi,
	&&\rk_{0x}=k\,\rn_{0x}=k\sin\vartheta_0\cos\varphi_0.
	\label{kx-kx0}
	\end{align}}

\begin{figure}
   { \begin{center}
        \includegraphics[scale=.27]{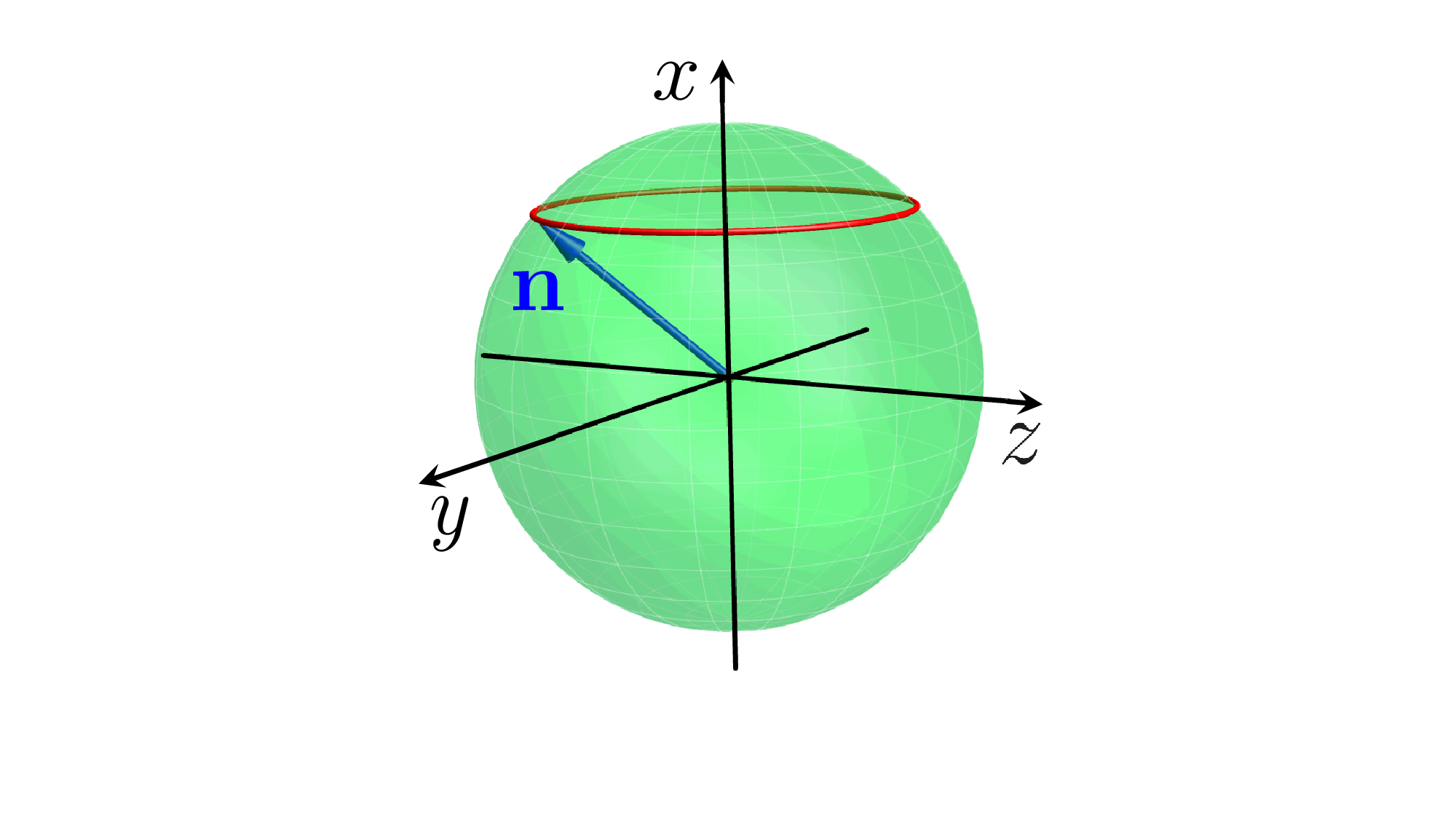}\vspace{-1cm}
        \caption{The unit sphere $S^2$ and the circle identifying the unit vectors $\bn$ with identical $x$ components $\rn_x\neq\pm1$. The blue arrow represents a generic example of such a unit vector.}
        \label{fig4}
    \end{center}}
    \end{figure}

{According to \eqref{overline-Rs} and \eqref{fz-def} -- \eqref{kx-kx0}, for fixed values of $k,d,\fz_1$, $\fz_2$, the function $Q(\vec\rk,\vec\rk_0)$, which determines the contribution of the point scatterers to the scattering amplitude, is a periodic function of $\sin\vartheta\cos\varphi$ which {equals} $\rn_x$. This means that the scattering effect due to the point scatterers along all unit vectors $\bn$ with the same $x$ component coincide. If we identify unit vectors with the points on the unit sphere $S^2$ given by $|\bfr|=1$, those with identical $x$ components $\rn_x$ correspond to the points on the intersection of $S^2$ and the plane {given by} $x=\rn_x$. As shown in Fig.~\ref{fig4}, this is a circle unless $\rn_x=\pm 1$ in which case $\bn=\pm\bfe_x:=(\pm 1,0,0)$.\footnote{{Due to the presence of the slab which prevents placing detectors on the $x$-$y$ plane, the latter choices for $\bn$ are forbidden.}} This phenomenon is a straightforward consequence of the axial symmetry of the point scatterer pair along the $x$ axis.}
	
{Next, suppose that $\fz_1$ and $\fz_2$ are so small that we can ignore the terms proportional to $\fz_1\fz_2$ in the right-hand side of \eqref{Q-a-zero}. Then in view of  \eqref{kx-kx0}, $Q(\vec\rk,\vec\rk_0)\approx 0$ if and only if there is a real number $\zeta\in[0,2\pi)$ and an integer $m$ such that 
	\begin{align}
	&\fz_2=\fz_1\,e^{i\zeta},
	&&\sin\vartheta\cos\varphi=\sin\vartheta_0\cos\varphi_0+\frac{\zeta+(2m+1)\pi}{dk}.
	\label{cancel}
	\end{align}
If these relations hold, the point scatterers' contributions to the scattering {amplitude} cancel and we find $\ff(\bn_0,\bn)\approx\mr\ff(\bn_0,\bn)$. {Specifically, for $\bn\neq\pm\bn_0$, we have $\ff(\bn_0,\bn)\approx 0$. This argument shows that by identifying the locations where $\bn\neq\pm\bn_0$ and the detectors register no scattered waves, one can determine the values of $\zeta$ and $d$ in terms of $k$ and $\sin\theta_0\cos\varphi_0$. If such locations does not exist, then one can conclude that $|\fz_1|\neq|\fz_2|$.}
	
{Since $|\sin\vartheta\cos\varphi-\sin\vartheta_0\cos\varphi_0|\leq 2$, \eqref{cancel}	
implies that one of the following scenarios {must hold}.
	\begin{enumerate}
	\item  $m=-1$ and $\zeta=\pi$: In this case $\fz_2=-\fz_1$ and 
	$\sin\vartheta\cos\varphi=\sin\vartheta_0\cos\varphi_0$, but $d$ is arbitrary.
	\item $|d|>\frac{\pi}{2k}=\frac{\lambda}{4}$.\nn
	\end{enumerate}
In particular, when $\fz_1=\fz_2$ and $|\fz_1|^2\ll 1$, $Q(\vec\rk,\vec\rk_0)\approx 0$ and $\ff(\bn_0,\bn)\approx\mr\ff(\bn_0,\bn)$ provided that $|d|>\frac{\lambda}{4}$ and 
	\be
	\sin\vartheta\cos\varphi=\sin\vartheta_0\cos\varphi_0+\frac{(2m+1)\pi}{dk}
	=\sin\vartheta_0\cos\varphi_0+\frac{\lambda}{d}\,\Big(m+\frac{1}{2}\Big).
	\label{dark}
	\ee
Notice that $|d|>\frac{\lambda}{4}$ is a necessary condition for this equation to hold, and for sufficiently large values of $\frac{|d|}{\lambda}$, there are finitely many integer $m$ fulfilling \eqref{dark}. Expressing this equation in the form, 
	\be
	\rn_x=\rn_{0x}+\frac{\lambda}{d}\,\Big(m+\frac{1}{2}\Big),
	\label{cancel-2}
	\ee
we observe that for each value of $\rn_{0x}$, there is a finite set of allowed values of $\rn_x$ for which $\ff(\bn_0,\bn)\approx\mr\ff(\bn_0,\bn)$.}

{For example consider the case of normal incidence with $\bn_0=\bfe_3:=(0,0,1)$. Then $\rn_{0x}=0$ and because $|\rn_{x}|\leq 1$, \eqref{cancel-2} implies $|m+\frac{1}{2}|\leq\frac{|d|}{\lambda}$. In Fig.~\ref{fig4} we identify the unit vectors $\bn$ with the points on the unit sphere {$S^2$} and plot the locus of points corresponding to those $\bn$ for which \eqref{cancel-2} holds and $d=2\lambda$. In this case, $m=-2,-1,0$ and $1$. These values of $m$ identify $\bn$ with the points lying on the intersection of the unit sphere and the planes given by $x=-\frac{3}{4}, -\frac{1}{4}, \frac{1}{4}$, and $\frac{3}{4}$. These are respectively the purple, blue, green, and red circles in Fig.~\ref{fig5}.}
	\begin{figure}
    {\begin{center}
        \includegraphics[scale=.25]{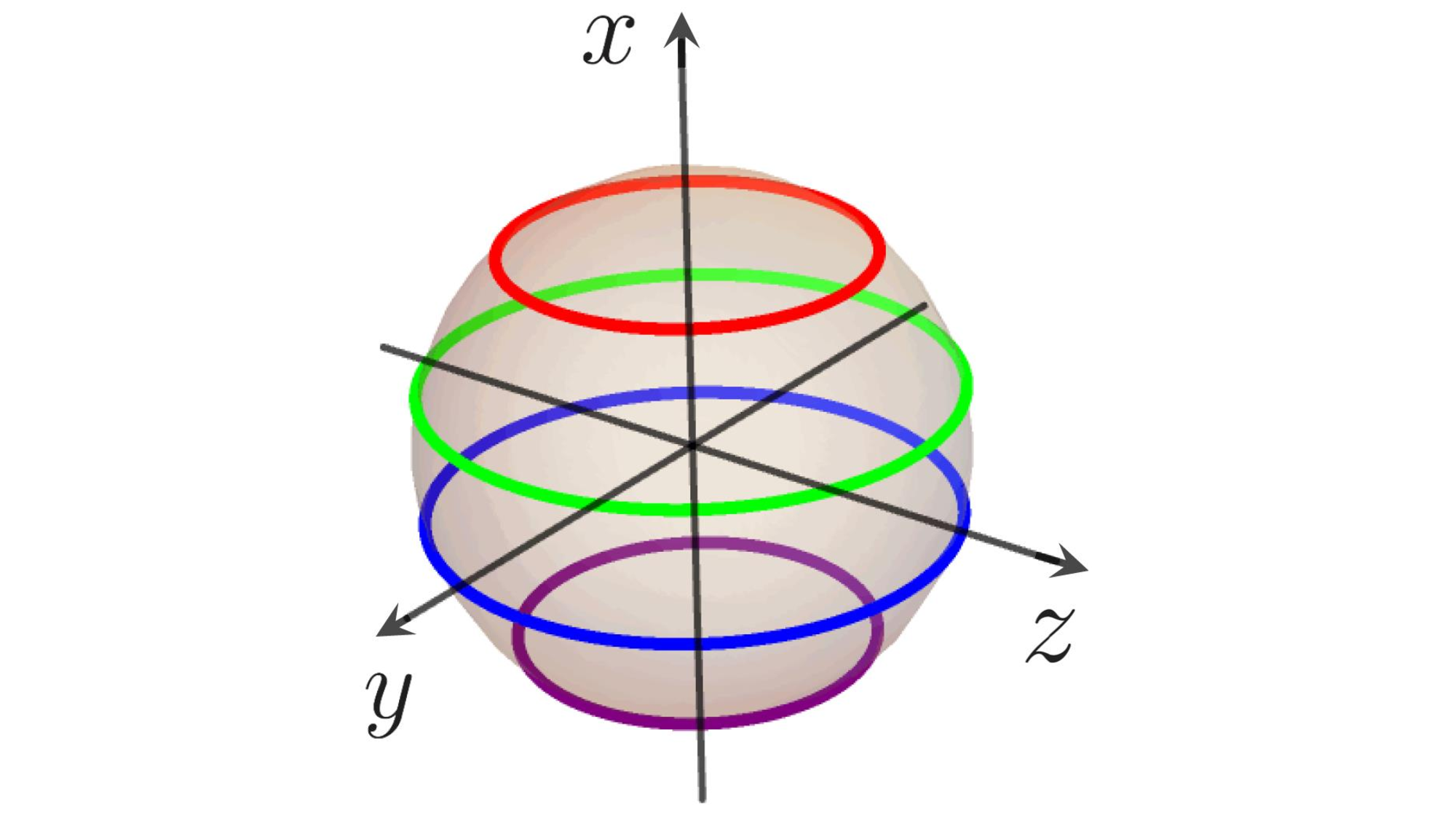}
        \caption{Graphical representation of the unit vectors $\bn$ along which the contributions of the point scatterers to the scattering amplitude cancel for $\fz_2=\fz_1$, $|\fz_1|\ll 1$, $d=2\lambda$, and $\bn_0=(0,0,1)=:\bfe_z$. {Because $\bn\neq\pm\bfe_z$, we have $\ff(\bfe_z,\bn)\approx 0$.} The pink sphere is the unit sphere $S^2$ that is centered at the origin. The circles plotted on $S^2$ represent the unit vectors $\bn$ fulfilling {$\ff(\bfe_z,\bn)\approx 0$}. For $d=2\lambda$, the integer $m$ appearing in \eqref{dark} take four values, namely $-2,-1,0$, and $1$. These correspond to the purple ($\rn_x=-\frac{3}{4}$), blue ($\rn_x=-\frac{1}{4}$), green ($\rn_x=\frac{1}{4}$), and red ($\rn_x=\frac{3}{4}$) curves, respectively.}
        \label{fig5}
    \end{center}}
    \end{figure}
	
If $\det(\bcA)$, which we denote by $D$ vanishes, the right-hand side of \eqref{Q-a-zero} and consequently $\ff(\bn_0,\bn)$ diverges. This marks a spectral singularity at which the point scatterers begin emitting coherence sound waves by amplifying the background noise. According to \eqref{fz-def}, this happens if and only if  
	\begin{align}
	\fz_2=\frac{(\overline{R}_1+1)\fz_1+i}{-(\overline{R}_1+1)+i\big\{[\overline{R}_2+\sinc(dk)]^2-
	(\overline{R}_1+1)^2\big\}\fz_1}.
	\label{SASER}
	\end{align}
For $dk\ll 1$, we can use \eqref{overline-Rs} to infer that $\overline{R}_2=\overline{R}_1+\cO(dk)^2$. We also have $\sinc(dk)=1+\cO(dk)^2$. In view of these relations, \eqref{Q-a-zero}, and \eqref{kx-kx0},  
	\begin{align}
	Q(\vec\rk,\vec \rk_0)
	&=\frac{\fz_1+\fz_2[1-idk(\sin\vartheta\cos\varphi-\sin\vartheta_0\cos\varphi_0)]}{k[1-i(\fz_1+\fz_2)(\overline{R}_1+1)]}+\cO(dk)^2.
	\label{Q-short}
	\end{align}
Moreover, the condition \eqref{SASER} for the emergence of a spectral singularity takes the form	
	\begin{align}
	\fz_1+\fz_2=-\frac{i}{\overline{R}_1+1}+\cO(dk)^2.
	\end{align}
In terms of $\check\beta_1$ and $\check\beta_2$, this relation reads
	\be
	\check\beta_1+\check\beta_2=-\frac{4\pi\,i}{k^3(1+\overline{R}_1)}+\cO(dk)^2.
	\ee
	
Next, consider the application of our general results to the scattering of sound waves with frequency 10 kHz by a 10~cm thick acoustic slab made of PDMS (polydimethylsiloxane) that is immersed in water, so that
	\begin{align}
	&\lambda=0.148~{\rm m},
	&&k= 42.45~{\rm m}^{-1},
	&&\ell=0.100~{\rm m},
	\label{spec-1}\\
	&\rho_0=1000~{\rm kg}/{\rm m}^3.
	&&c_0=1480~{\rm m}/{\rm s},
	&&\rho_s=920~{\rm kg}/{\rm m}^3,
	\label{spec-2}\\
	&c_s=1077~{\rm m}/{\rm s},
	&&\alpha_s=0.920,
	&&\fn_s=1.374,
	\label{spec-3}\\
	&\kappa_0=4.565\times 10^{-10}~{\rm Pa}^{-1},
	&&\kappa_s=9.371\times 10^{-10}~{\rm Pa}^{-1},
	\label{spec-4}
	\end{align}
where $\lambda$ stands for the incident wavelength (in water), and $\rho_s$ and $c_s$ respectively denote the density and speed of sound in the PDMS slab \cite{Li-2023}. 

Suppose that the slab is in the configuration depicted in Fig.~\ref{fig3}, and a pair of identical air-filled macro-spheres of diameter 2 mm made of soft outer shells are placed on the face of the slab corresponding to the $x$-$y$ plane. The shells' densities are elevated by addition of high-density nanoparticles so that the effective density of the micro-spheres is close to that of the slab but their thickness and softness is such that their effective compressibility ($\kappa_{\rm eff}$) is much larger than that of the slab. For example, {assume} that
	\be
	\kappa_{\rm eff}=\frac{\kappa_{\rm air}}{10}=7.083\times 10^{-7}~{\rm Pa}^{-1}.	\label{kappa-eff=}
	\ee
Since the incident wavelength is much larger than the diameter of the macro-spheres, we can view the latter as point scatterers. {Given} that their centers are located at $\bfa_1=(0,0,0)$ and $\bfa_2=(d,0,0)$, and noting that in view of \eqref{spec-4} and \eqref{kappa-eff=}, $\kappa_{\rm eff}=1551 \kappa_0=756 \kappa_s$, we can justify modeling the compressibility of the system using \eqref{s6-kappa} with $n=2$.

{To determine the values of the coefficients $\check\kappa_1$ and $\check\kappa_2$ of \eqref{s6-kappa}, we identify them with \linebreak $(\kappa_{\rm eff}-\overline\kappa)V_{\rm p.s.}$, where  $\overline\kappa:=\frac{1}{2}(\kappa_0+\kappa_s)$ and $V_{\rm p.s.}$ is the volume of the macro-spheres. In view of \eqref{spec-2} and the fact that the radii of point scatterers equal $1$~mm, this gives $\check\kappa_1=\check\kappa_2=2.96\times 10^{-15}~{\rm Pa}^{-1}$. Substituting this in the last equation in \eqref{mr-alpha-def} and making use of \eqref{fz-i=} and \eqref{spec-1}, we obtain
	\begin{align}
	&\check\beta_1=\check\beta_2=6.49\times 10^{-6}~{\rm m}^3,
	&&\fz_1=\fz_2=3.95\times 10^{-2}.
	\label{betas=}
	\end{align}
Because $|\fz_1|^2<1.6\times 10^{-3}$, the above results on the vanishing of $Q(\vec\rk,\vec\rk_0)$
apply.}

Figures~\ref{fig6} and \ref{fig7} show the plots of the differential cross section $|\ff(\bn_0,\bn)|^2$ as functions of $\vartheta$ and $\varphi$ for a normally incident wave with $\vartheta_0=0$. To obtain these plots, we have used the exact expressions for $\overline R_1$, $\overline R_2$, and $Q(\vec\rk,\vec\rk_0)$, i.e., Eqs.~\eqref{overline-Rs} and \eqref{Q-a-zero}. This is because for the chosen values of $d$ and $k$, $(dk)^2>0.16$.
\begin{figure}
        {\begin{center}
        \includegraphics[scale=.55]{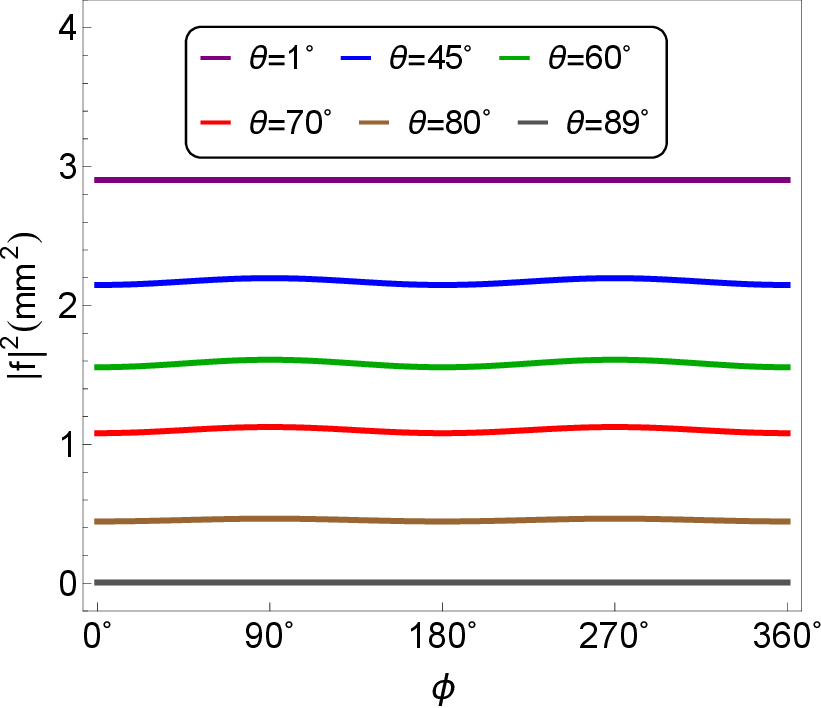}~~\includegraphics[scale=.57]{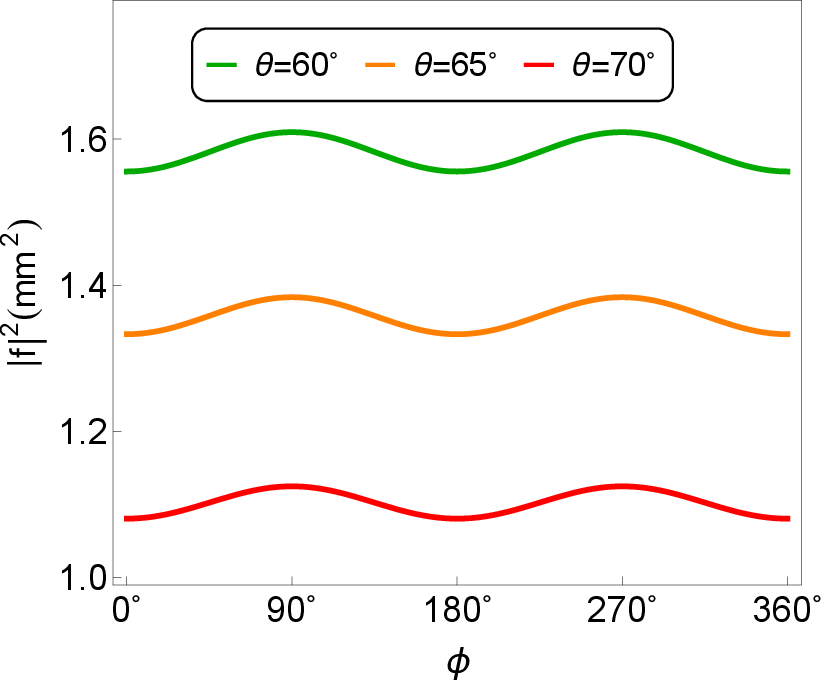}
       \caption{Graphs of the differential cross section $|\ff|^2$ as a function of $\varphi$ for for $\vartheta_0=0$ and different values of $\vartheta$. As $\varphi$ changes, $|\ff|^2$ oscillates with an amplitude less than $0.05$.}
        \label{fig6}
        \end{center}}
        \end{figure}
        \begin{figure}
        {\begin{center}
        \includegraphics[scale=.6]{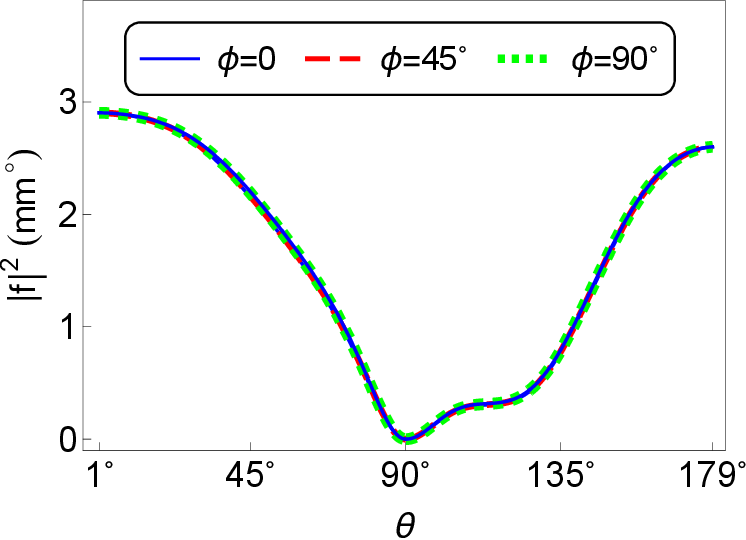}
       \caption{Graph of the differential cross section $|\ff|^2$ as a function of $\vartheta$ for $\vartheta_0=0$ and different values of $\varphi$. The graphs for different $\varphi$ appear to coincide, because as shown in Fig.~\ref{fig6}, the amplitudes of the variations of the cross section due to changes in $\varphi$ are two orders of magnitude smaller than those arising from changes in $\vartheta$.}
        \label{fig7}
        \end{center}}
        \end{figure}
        
In the absence of the homogeneous slab, $\mr R(\vartheta)=0$, $\mr T(\vartheta)=1$, and \eqref{f-a=zero} reduces to $\ff(\bn_0,\bn)=Q(\vec\rk,\vec\rk_0)$. Fig.~\ref{fig8} shows the the graphs of the differential cross section as a function of $\vartheta$ for $\vartheta_0=0$ and different values of $\varphi$ for the point scatterers in the absence of the slab. The dependence of the differential cross section on $\varphi$ is more pronounced in this figure. The differential cross section attains its maximum value (3.45~mm$^2$) for $\varphi=90^\circ$, i.e., when $\bn$ lies on the $y$-$z$ plane. The presence of the slab is responsible for the drastic difference in the behavior of the differential cross sections depicted in Figs.~\ref{fig7} and \ref{fig8}. In particular, it causes the differential cross section to vanish for $\vartheta\approx 90^\circ$, i.e., for grazing angles. While in the absence of the slab the differential cross section has a value larger than 3.30~mm$^2$ for $\vartheta\approx 90^\circ$.
	  \begin{figure}
        {\begin{center}
        \includegraphics[scale=.6]{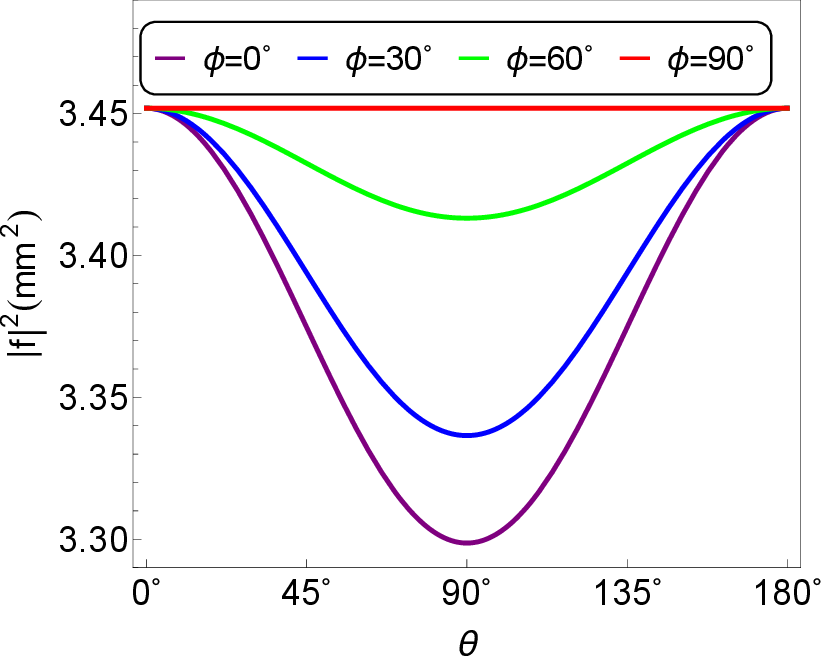}
       \caption{Graph of the differential cross section $|\ff|^2$ as a function of $\vartheta$ for $\vartheta_0=0$ and different values of $\varphi$ in the absence of the homogeneous slab. The amplitudes of the variations of the cross section due to changes in $\varphi$ are comparable with those arising from changes in $\vartheta$. The maximum value of the differential cross section is 3.45~mm$^2$. It is attained for $\varphi=90^\circ$.}
        \label{fig8}
        \end{center}}
        \end{figure}

\section{Concluding remarks}

Acoustic scattering is a well-developed area of research with a long history. In this article, we propose an alternative approach to its treatment in which the scattering features of an inhomogeneous compressible carrier medium is derived from the dynamics of an effective non-unitary quantum system. This extends our earlier results on the dynamical formulation of stationary scattering for quantum and electromagnetic scattering to acoustic scattering in three dimensions. It also highlights the importance of the study of quantum dynamics generated by non-Hermitian Hamiltonians {in the study of wave propagation and scattering.}

Our approach allows for an alternative proof of the acoustic reciprocity theorem and leads to explicit analytic formulas for the leading- and next-to-leading-order terms in the low-frequency series expansion of the acoustic scattering amplitude. Another of its important applications is in the study of delta-function point scatterers {whose standard treatment involves regularization of the emergent singularities and their removal via a coupling-constant renormalization \cite{ap-2022,jackiw,Mitra}. We have used it to obtain an exact analytic solution of the scattering problem for an acoustic slab including a planar array of delta-function point scatterers. The resulting formula for the scattering amplitude relates the locations of the point scatterers to the scattering data. Therefore it offers a direct inverse scattering prescription.}

In this article, we have confined our attention to the theoretical aspects of the dynamical formulation of stationary acoustic scattering and its prospects for performing analytic treatments of specific scattering problems. {An entirely open area of research is to explore its utility in numerical acoustic scattering. Another direction of future research is the study of the mathematical properties of its main ingredient, namely the fundamental transfer matrix.}

\subsection*{Acknowledgements} 
This work has been supported by the Scientific and Technological Research Council of T\"urkiye (T\"UB\.{I}TAK) in the framework of the project 123F180 and by Turkish Academy of Sciences (T\"UBA).  

\subsection*{{Appendix~A: Consequences of translations along the $z$ axis}}

Consider a compressible fluid $F$ determined by the mass density $\rho$ and compressibility $\kappa$, and let $\ra$ be a real parameter with the dimension of length. In this appendix we derive the transformation rule for the fundamental transfer matrix $\widehat\bM$ and the scattering amplitude $\ff$ of $F$ under a space translation of the form, 
	\be
	(\vec r,z)=\bfr\to \overline\bfr:=(\vec r,z-\ra), 
	\label{AppA-1}
	\ee
which maps $F$ to the fluid $\overline F$ with mass density $\overline\rho$ and compressibility $\overline\kappa$ given by $\overline\rho(\bfr):=\rho(\overline\bfr)$ and $\overline\kappa(\bfr):=\kappa(\overline\bfr)$. 

Suppose that $\bp_\pm:=(\vec p,\pm p_z)\in\R^3$ and $f_\pm:\R^3\to\C$ are the scalar functions defined by $f_\pm(\bfr):=e^{i\bp_\pm\cdot\bfr}=e^{i\vec p\cdot\vec r}e^{\pm i p_z z}$. Then, under the translation \eqref{AppA-1}, $f_\pm$ transform according to
	\be
	f_\pm(\bfr)\to \overline f_\pm(\bfr):=f_\pm(\overline{\bfr})=
	f_\pm(\vec r,z-\ra)=e^{i\vec p\cdot\vec r}e^{\pm i p_z (z-\ra)}=
	e^{\mp i \ra p_z}f_\pm(\vec r,z).\nn
	\ee
Applying this transformation rule to the asymptotic expression \eqref{asymp} for the bounded solutions of the Bergmann equation \eqref{bergmann-eq}, we find {that} the effect of the translation \eqref{AppA-1} on the latter is equivalent to transforming the coefficient functions $A_\pm$ and $B_\pm$ as follows:
	\begin{align}
	&A_\pm(\vec p)\to \overline A_\pm(\vec p):=e^{-i\ra\varpi(\vec p\,)}A_\pm(\vec p),
	&B_\pm(\vec p)\to \overline B_\pm(\vec p):=e^{i\ra\varpi(\vec p\,)}B_\pm(\vec p).
	\nn
	\end{align}
By virtue of (\ref{hat-varpi}), we can express these relations in the form
	 \begin{align}
	&A_\pm \to \overline A_\pm:=e^{-i\ra\widehat\varpi }A_\pm,
	&&B_\pm\to \overline B:=e^{i\ra\widehat\varpi}B_\pm.
	\label{AppA-2}
	\end{align}
	
Next, consider the defining relation for the the fundamental transfer matrix $\widehat{\overline\bM}$ of $\overline F$, namely
	\be
	\left[\begin{array}{c}
	\overline A_+\\
	\overline B_+\end{array}\right]=\widehat{\overline\bM}\left[\begin{array}{c}
	\overline A_-\\
	\overline B_-\end{array}\right].
	\label{translated-M}
	\ee
In view of \eqref{M-def}, \eqref{AppA-2}, and \eqref{translated-M},  
	\be
	\widehat{\overline\bM}=e^{-ia\widehat\varpi\bsigma_3}\widehat\bM\,
	e^{ia\widehat\varpi\bsigma_3}.
	\label{M-trans}
	\ee
Expressing this relation in terms of the entries of $\widehat{\overline\bM}$ and $\widehat\bM$, we have
	\begin{align}
	&
	\widehat{\overline M}_{11}=e^{-i\ra\widehat\varpi}\widehat M_{11}e^{i\ra\widehat\varpi},
	&&
	\widehat{\overline M}_{12}=e^{-i\ra\widehat\varpi}\widehat M_{12}e^{-i\ra\widehat\varpi},
	\nn\\
	&
	\widehat{\overline M}_{21}=e^{i\ra\widehat\varpi}\widehat M_{21}e^{i\ra\widehat\varpi},
	&&
	\widehat{\overline M}_{22}=e^{i\ra\widehat\varpi}\widehat M_{22}e^{-i\ra\widehat\varpi}.
	\nn
	\end{align}
These in turn imply
	\begin{align}
	&
	\widehat{\overline M}^{-1}_{22}=e^{i\ra\widehat\varpi}\widehat M_{22}^{-1}e^{-i\ra\widehat\varpi},
	\label{Mab-trans-22}\\
	&
	\widehat{\overline M}_{22}^{-1}\widehat{\overline M}_{21}=
	e^{i\ra\widehat\varpi}\widehat M_{22}^{-1}\widehat M_{21}e^{i\ra\widehat\varpi},
	\label{Mab-trans-21}\\
	&
	\widehat{\overline M}_{12}\widehat{\overline M}_{22}^{-1}=
	e^{-i\ra\widehat\varpi}\widehat M_{12}\widehat M_{22}^{-1}e^{-i\ra\widehat\varpi}.
	\label{Mab-trans-12}
	\end{align}

The scattering amplitude $\overline\ff$ for $\overline F$ is given by \eqref{ff=defect} -- \eqref{fmm=} with $\widehat M_{ab}$ replaced with $\widehat{\overline{M}}_{ab}$. In view of \eqref{Mab-trans-22} -- \eqref{Mab-trans-12}, this leads to the following transformation rule for the scattering amplitude.
	\be
	\ff(\bn_0,\bn)\to\overline\ff(\bn_0,\bn)= \left\{\begin{array}{ccc}
	\overline\ff_{+,+}&\for&\rn_{0z}>0~{\rm and}~\rn_z>0,\\
	\overline\ff_{+,-}&\for&\rn_{0z}>0~{\rm and}~\rn_z<0,\\
	\overline\ff_{-,+}&\for&\rn_{0z}<0~{\rm and}~\rn_z>0,\\
	\overline\ff_{-,-}&\for&\rn_{0z}<0~{\rm and}~\rn_z<0,
	\end{array}\right.
	\label{ff=defect-2}
	\ee
where
	\begin{align}
	&\overline\ff_{\pm,\pm}:=e^{\pm i\ra[\varpi(\vec\rk_0)-\varpi(\vec\rk)]}\,\ff_{\pm,\pm},
	&&\overline\ff_{\pm,\mp}:=e^{\pm i\ra[\varpi(\vec\rk_0)+\varpi(\vec\rk)]}\,\ff_{\pm,\mp}.
	\label{f-trans-1}
	\end{align}

\subsection*{Appendix~B: Solution of Eq.~\eqref{int-eq}}

To solve \eqref{int-eq}, first we use \eqref{M=M+Ms}, \eqref{mr-RRT}, and \eqref{hat-mr-T} to show that
	\be
	\widehat M_{22}=\widehat{\mr M}_{21}\widehat\sM_{12}+
	\widehat{\mr M}_{22}(\widehat I+\widehat\sM_{22})
	=\widehat{\mr T}\,^{-1}\big(\widehat I-\widehat{\mr R}\,^l\widehat\sM_{12}+
	\widehat\sM_{22}\big),
	\label{App-B-eq1}
	\ee
where $\widehat{\mr M}_{ab}$ and $\widehat{\sM}_{ab}$ are respectively the entries of $\widehat{\mr\bM}$ and $\widehat{\bsM}$, and $\widehat{\mr R}\,^l:=\mr R^l(\widehat{\vec p}\,)$. Recalling that $\mr u_{ab}$ are the entries of $\bcU(\vec p,0,-\infty)$ and using \eqref{cK=}, \eqref{s6-bsM=}, \eqref{mr-U-inv}, \eqref{cX-def} -- \eqref{hat-mr-T}, we have
	\begin{align}
	& \widehat\sM_{12}=\frac{ik^2}{2}\,\widehat\cX\,
	\widehat\Pi_k\widehat\fB\,\widehat\Pi_k\widehat\varpi^{-1}\widehat\cX,
	\label{sM-12}\\
	& \widehat\sM_{22}=-\frac{ik^2}{2}\,\widehat\cY\,
	\widehat\Pi_k\widehat\fB\,\widehat\Pi_k\widehat\varpi^{-1}\widehat\cX,
	\label{sM-22}
	\end{align}
where $\widehat\cY:=\cY(\widehat{\vec p}\,)$. In view of \eqref{beta-prime-def}, \eqref{cZ-def}, and \eqref{App-B-eq1} -- \eqref{sM-22}, we can express \eqref{int-eq} in the form:
	\begin{align}
	|\chi\kt&=\frac{ik^2}{2}\,\widehat\cZ\,
	\widehat\Pi_k\widehat\fB\,\widehat\Pi_k\widehat\varpi^{-1}\widehat\cX|\chi\kt+
	\widehat{\mr T}|\phi\kt\nn\\
	&=i\pi k^2\, \sum_{i=1}^n\check\beta_i\br\vec\ra_i|\cF^{-1}|\chi'\kt\:
	\widehat\cZ\,\widehat\Pi_k|\vec\ra_i\kt+\widehat{\mr T}|\phi\kt,
	\label{App-B-eq2}
	\end{align}
where
	\be
	|\chi'\kt:=\widehat\varpi^{-1}\widehat\cX|\chi\kt,
	\label{chi-prime}
	\ee
and we have made use of the identity, $\widehat\Pi_k|\chi'\kt=|\chi'\kt$, which follows from $|\chi\kt\in\sF_k$ and the fact that $\widehat\varpi^{-1}$ and $\widehat\cX$ commute with $\widehat\Pi_k$.
	
Next, we apply $\br\vec\ra_j|\cF^{-1}\widehat\varpi^{-1}\widehat\cX$ to both sides of \eqref{App-B-eq2} and use \eqref{chi-prime} to express the resulting equation in the form:
	\begin{align}
	\sum_{i=1}^n \cA_{ji}\,\br\vec\ra_i|\cF^{-1}|\chi'\kt=\br\vec\ra_j|\cF^{-1}
	\widehat\varpi^{-1}\widehat\cX\,\widehat{\mr T}|\phi\kt,
	\label{App-B-eq4}
	\end{align}
where $\cA_{ij}$ are given by \eqref{cA-ij=}.  \eqref{App-B-eq4} is a linear system of $n$ equations for $\br\vec\ra_i|\cF^{-1}|\chi'\kt$. It has the following unique solution provided that its matrix of coefficients, namely $\bcA:=[\cA_{ij}]$, is invertible. 
	\be
	\br\vec\ra_i|\cF^{-1}|\chi'\kt=\sum_{j=1}^n\cA_{ij}^{-1}
	\br\vec\ra_j|\cF^{-1}
	\widehat\varpi^{-1}\widehat\cX\,\widehat{\mr T}|\phi\kt,
	\label{App-B-eq5}
	\ee
where $\cA_{ij}^{-1}$ are the entries of the inverse of $\bcA$. Substituting \eqref{App-B-eq5} in \eqref{App-B-eq2}, we find \eqref{M22-inv=}.

\subsection*{Appendix~C: Derivation of Equations \eqref{fpp-0} -- \eqref{fmm-0}}

In this appendix we provide details of the calculation of $\ff_{\pm,\pm}$ and $\ff_{\mp,\pm}$.

First, we draw attention to the fact that, according to \eqref{varpi=}, \eqref{mr-V-63}, \eqref{mr-H-63}, and \eqref{mr-U-def}, $\varpi(\vec p\,)$, $\mr\sV(z,\vec p\,)$, and consequently $\mr\bcH(z,\vec p\,)$ and $\mr\bcU(\vec p,z,z_0)$ are functions of $|\vec p\,|^2$. In view of \eqref{mr-M-def-62}, \eqref{mr-RRT}, and \eqref{cX-def} -- \eqref{hat-mr-T}, 
the same applies to $\mr\bM_{ab}(\vec p\,)$, $\mr R^{l/r}(\vec p\,)$, $\mr T(\vec p\,)$, $\cX(\vec p\,)$, $\cY(\vec p\,)$, and $\cZ(\vec p\,)$. In particular,
	\begin{align}
	&\mr R^{l/r}(-\vec p\,)=\mr R^{l/r}(\vec p\,), 
	&&\mr T(-\vec p\,)=\mr T(\vec p\,), 
	&&\cX(-\vec p\,)=\cX(\vec p\,),
	\label{id-501}\\ 
	&\cY(-\vec p\,)=\cY(\vec p\,), 
	&&\cZ(-\vec p\,)=\cZ(\vec p\,).
	\label{id-502}
	\end{align}
We can use these equations together with \eqref{cA-ij=} and \eqref{alpha-ij} to infer that
	\begin{align}
	&\alpha_{ij}=\alpha_{ji},
	&&\check\beta_i\cA_{ij}=\check\beta_j\cA_{ji},
	&&\check\beta_j\cA_{ji}^{-1}=\check\beta_iA_{ij}^{-1}.
	\label{sym}
	\end{align}
By virtue of the latter identity and \eqref{mr-fm},  
	\begin{align}
	Q(-\vec\rk_0,-\vec\rk)&=\frac{k^2}{4\pi}\sum_{i,j=1}^n\check\beta_i\,\cA_{ij}^{-1}
	e^{-i(\vec\ra_j\cdot\vec\rk-\vec\ra_i\cdot\vec\rk_0)}\nn\\
	&=\frac{k^2}{4\pi}\sum_{i,j=1}^n\check\beta_j\,\cA_{ji}^{-1}
	e^{-i(\vec\ra_i\cdot\vec\rk-\vec\ra_j\cdot\vec\rk_0)}\nn\\
	&=\frac{k^2}{4\pi}\sum_{i,j=1}^n\check\beta_i\,\cA_{ij}^{-1}
	e^{-i(\vec\ra_i\cdot\vec\rk-\vec\ra_j\cdot\vec\rk_0)}\nn\\
	&=Q(\vec\rk,\vec\rk_0).
	\label{id-300}
	\end{align}	
	
Next, we set $|\phi\kt=|\vec\rk_0\kt$ in \eqref{M22-inv=}, apply  $\br\vec p\,|$ from the left to both sides of the resulting equation, and use \eqref{gamma-j}, \eqref{b-j}, and  \eqref{mr-fm} to show that, for $\vec p\in\sD_k$,
	\be
	\br\vec p\,|\widehat M_{22}^{-1}|\vec\rk_0\kt=
	\mr T(\vec\rk_0)\big[\delta(\vec p-\vec\rk_0)+\tfrac{i}{2\pi}\,\varpi(\vec\rk_0)^{-1} Q(\vec p,\vec\rk_0)\cX(\vec\rk_0)\cZ(\vec p\,)\big].
	\label{AppC-eq146}
	\ee
Equations \eqref{fpp-0} and \eqref{fmm-0} follow from \eqref{fpp=},  \eqref{fmm=},  \eqref{id-501}, \eqref{id-502}, \eqref{id-300}, and \eqref{AppC-eq146}.

To calculate $\ff_{\pm,\mp}$, first we use   \eqref{cK=}, \eqref{M=M+Ms}, \eqref{s6-bsM=}, \eqref{mr-RRT}, \eqref{mr-U-inv}, and \eqref{cX-def} -- \eqref{hat-mr-T} to establish
	\begin{align}
	&\widehat M_{12}=\widehat{\mr M}_{11}\widehat\sM_{12}+
	\widehat{\mr M}_{12} \big(\widehat\sM_{22}+\widehat I\big)
	=\widehat{\mr T}\,^{-1}\big[\big(\widehat{\mr T}\,^2-\widehat{\mr R}\,^l
	\widehat{\mr R}\,^r\big)\widehat\sM_{12}+\widehat{\mr R}\,^r\big(\widehat\sM_{22}+\widehat I\big)\big],
	\label{App-C-eq11}\\
	&\widehat M_{21}=\widehat{\mr M}_{21}(\widehat I+\widehat\sM_{11})+
	\widehat{\mr M}_{22} \widehat\sM_{21}
	=\widehat{\mr T}\,^{-1}\big[\widehat\sM_{21}-\widehat{\mr R}\,^l(
	\widehat I+\widehat\sM_{11})\big],
	\label{App-C-eq1}\\
	& \widehat\sM_{11}=\frac{ik^2}{2}\,\widehat\cX\,
	\widehat\Pi_k\widehat\fB\,\widehat\Pi_k\widehat\varpi^{-1}\widehat\cY,
	\quad\quad
	\widehat\sM_{21}=-\frac{ik^2}{2}\,\widehat\cY\,
	\widehat\Pi_k\widehat\fB\,\widehat\Pi_k\widehat\varpi^{-1}\widehat\cY.
	\label{sM-11-21}
	\end{align}
Substituting \eqref{sM-12}, \eqref{sM-22}, and \eqref{sM-11-21} in \eqref{App-C-eq11} and \eqref{App-C-eq1} and making use of \eqref{cZ-def}, we have
	\begin{align}
	\widehat M_{12}&=\widehat{\mr T}\,^{-1}\big[\widehat{\mr R}\,^r+
	\frac{ik^2}{2}
	\big(\widehat{\mr T}\,^2\widehat\cX-\widehat{\mr R}\,^r\widehat\cZ\big)
	\widehat\Pi_k\widehat\fB\widehat\Pi_k\widehat\varpi^{-1}\widehat\cX\big],
	\label{App-C-eq21}\\
	\widehat M_{21}&=-\widehat{\mr T}\,^{-1}\big(\widehat{\mr R}\,^l+
	\frac{ik^2}{2}\,\widehat\cZ\,
	\widehat\Pi_k\widehat\fB\,\widehat\Pi_k\widehat\varpi^{-1}\widehat\cY\big).
	\label{App-C-eq2}
	\end{align}
	
Next, let $\vec p\in\sD_k$ and $|\phi\kt:=\widehat M_{21}|\vec\rk_0\kt$. Then $|\phi\kt\in\sF_k$, and in view of \eqref{beta-prime-def}, \eqref{M22-inv=}, and \eqref{App-C-eq2}, we have
	\begin{align}
	&\begin{aligned}
	\br\vec\rk\,|\widehat M_{22}^{-1}\widehat M_{21}|\vec\rk_0\kt&=
	\br\vec\rk\,|\widehat M_{22}^{-1}|\phi\kt={\mr T}(\vec\rk\,)\phi(\vec\rk)+
	\frac{i k^2}{2}\,\cZ(\vec\rk\,)
	\sum_{i,j=1}^n\check\beta_i\, \cA_{ij}^{-1}\Gamma_j\, e^{-i\vec\ra_i\cdot\vec\rk},
	\end{aligned}
	\label{App-C-eq3}\\
	&\begin{aligned}
	\phi(\vec p\,)&=\br\vec p\,|\phi\kt=-\frac{1}{\mr T(\vec p\,)}
	\Big[\mr R^l(\vec\rk_0)\delta(\vec p-\vec\rk_0)+\frac{ik^2\cY(\vec\rk_0)}{2\,\varpi(\vec\rk_0)}\,
	\cZ(\vec p\,)\,\br p|\widehat\fB|\vec\rk_0\kt\Big]\\
	&=-\frac{1}{\mr T(\vec p\,)}
	\Big[\mr R^l(\vec\rk_0)\delta(\vec p-\vec\rk_0)+
	\frac{ik^2\cY(\vec\rk_0)}{8\pi^2\varpi(\vec\rk_0)}\,
	\cZ(\vec p\,)\,\sum_{i=1}^n\check\beta_i e^{-i\vec\ra_i(\vec p-\vec\rk_0)}\Big].
	\end{aligned}
	\label{App-C-eq4}
	\end{align}
Substituting \eqref{App-C-eq4} in \eqref{gamma-j}, we obtain
	\be
	\Gamma_j=-\frac{1}{4\pi^2\varpi(\vec\rk_0)}\left[ e^{i\vec\ra_j\cdot\vec\rk_0}\cX(\rk_0)\mr R^l(\vec\rk_0) 
	+i\cY(\vec\rk_0)\sum_{l=1}^n\alpha_{jl} \check\beta_l e^{i\vec\ra_l\cdot\vec\rk_0}\right],
	\label{Gamma=0}
	\ee
where $\alpha_{jl}$ is defined by \eqref{alpha-ij}. According to \eqref{cA-ij=}, the latter satisfies 
	\be
	\sum_{j=1}^n\cA_{ij}^{-1}\alpha_{jl}\check\beta_l=
	i\sum_{j=1}^n\cA_{ij}^{-1}(\cA_{jl}-\delta_{jl})=i(\delta_{il}-\cA_{il}^{-1}).
	\label{id-200}
	\ee
Equation~\eqref{fpm-0} follows from \eqref{fpm=}, \eqref{App-C-eq3} --  \eqref{id-200}. This completes our calculation of $\ff_{+,-}$.

To find an explicit expression for $\ff_{-,+}$, first we use \eqref{beta-prime-def}, 
and \eqref{App-C-eq21} to show that
	\begin{align}
	\br\vec\rk\,|\widehat M_{12}|\vec p\,\kt
	&=\mr T(\vec\rk\,)^{-1}\bigg[
	\mr R\,^r(\vec\rk\,)\,\delta(\vec\rk-\vec p\,)+\frac{i k^2 \cW(\vec\rk\,)\cX(\vec p\,)
	}{8\pi^2\varpi(\vec p\,)}\sum_{i=1}^n\check\beta_i\, e^{-i\vec\ra_i\cdot(\vec\rk-\vec p\,)}
	\bigg],
	\label{App-C-41}
	\end{align}
where
	\be
	\cW(\vec p):=\mr T(\vec p\,)^2\cX(\vec p\,)-\mr R^r(\vec p\,)\cZ(\vec p\,).
	\label{cW-def}
	\ee
Substituting \eqref{AppC-eq146} and \eqref{App-C-41} in the right-hand side of the identity,
	\be
	\br\vec\rk\,|\widehat M_{12}\widehat M_{22}^{-1}|\vec\rk_0\kt=
	\int_{\sD_k} d^2p\: \br\vec\rk\,|\widehat M_{12}|\vec p\,\kt\,\br\vec p|\widehat M_{22}^{-1}|\vec\rk_0\kt,\nn
	\ee	
and using \eqref{cZ-def}, \eqref{fmp=}, \eqref{id-200}, and \eqref{cW-def}, we obtain \eqref{fmp-0}.

\section*{Appendix~D: Calculation of $\mr\bcU(\vec p,z,z_0)$ for a homogenous slab}
 
 Consider the scattering of the homogeneous acoustic slab $\mr\cS$ described in Subsec.~6.4. According to \eqref{s64-z2}, $\mr\alpha(\vec r,z)=\mr\beta(\vec r,a)=1$ for $z<-\ra$ and $z>\ell-\ra$. This relation together with \eqref{mr-V-63} and \eqref{mr-H-63} imply
	\begin{align}
	\mr\bcH({\vec p},0)&=\frac{1}{2}\,
	\Big\{\big[(1-\alpha_s^{-1})\varpi(\vec p\,)+k^2(\alpha_s^{-1}-\beta_s)\varpi(\vec p)^{-1}\big]\bcK+(1- \alpha_s)\varpi(\vec p\,) \bcK^T\Big\},\\
	\mr\bcH({\vec p},z)&=\chi(z)\,e^{-iz\varpi(\vec p\,)\bsigma_3} \,
	\mr\bcH({\vec p},0)\, e^{iz\varpi(\vec p\,)\bsigma_3},
	\label{mr-H-6s-a}
	\end{align}
where
	\begin{align}
	&\chi(z):=\left\{\begin{array}{cc}
	1 &\for~-\ra\leq z\leq\ell-\ra,\\
	0 &\mbox{otherwise}.\end{array}\right.\nn
	\end{align}

Following, the treatment of homogeneous electromagnetic slabs in two-dimensions that is offered in Ref.~\cite{ptep-2026b}, we can use \eqref{mr-H-6s-a} to show that
	\be
	\mr\bcU(\vec p,z,z_0)=e^{-i z\varpi(\vec p\,)\bsigma_3}\,e^{-i(z-z_0)\mr\bH(\vec p\,)}\,
	e^{i z_0\varpi(\vec p\,)\bsigma_3}~~~\for~~-\ra\leq z_0\leq z\leq\ell-\ra,
	\label{mr-U=AppD}
	\ee
where
	\begin{align}
	\mr\bH(\vec p\,)&:= \mr\bcH(\vec p,0)-\varpi(\vec p\,)\bsigma_3=
	\varpi(\vec p\,)\left[\begin{array}{cc}
	- \fm_{+}(\vec p\,)&- \fm_{-}(\vec p\,) \\  
	\fm_{-}(\vec p\,) & \fm_{+}(\vec p\,)
	\end{array}\right]\nn\\
	&=\varpi(\vec p\,)\tilde\fn(\vec p\,)\bG(\vec p\,)\bsigma_3\bG(\vec p\,)^{-1},
	\label{mr-bH=}\\[3pt]
	\bG(\vec p\,)&:=\left[\begin{array}{cc}
	\alpha_s-\tilde\fn(\vec p\,) & \alpha_s+\tilde\fn(\vec p\,) \\
	 \alpha_s+\tilde\fn(\vec p\,) &  \alpha_s-\tilde\fn(\vec p\,)\end{array}\right],
	 \label{G-def}
	\end{align}
and $\fm_\pm$ and $\tilde\fn$ are defined by \eqref{m-tn-def}. According to  \eqref{mr-bH=}, 
	\[e^{-i(z-z_0)\mr\bH(\vec p\,)}=\bG(\vec p\,)\,
e^{-i(z-z_0)\tilde\fn(\vec p\,)\varpi(\vec p\,)\bsigma_3}\,\bG(\vec p\,)^{-1}.\]
Substituting \eqref{G-def} in this relation and using the result to evaluate the right-hand side of \eqref{mr-U=AppD}, we are led to Eqs.~\eqref{cU11=} -- \eqref{cU22=} for the entries $\mr\cU_{ab}(\vec p,z,z_0)$ of $\mr\bcU(\vec p,z,z_0)$.

\ed